\documentclass[10pt,twocolumn,twoside]{IEEEtran}

\usepackage{mathrsfs}

\newcommand{\norm}[1]{\left\lVert#1\right\rVert}

\usepackage{cite}
\usepackage[cmex10]{amsmath}
\usepackage{amsthm, amssymb}
\usepackage{euscript}

\theoremstyle{definition}

\usepackage{setspace}
 \usepackage{verbatim}
\usepackage{epsfig}
\usepackage{graphicx}
\usepackage[caption=false,font=footnotesize]{subfi
g} % IEEEtran does not allow loading 'caption'
\usepackage{epstopdf} % to import .eps figures
\usepackage{arydshln}
\usepackage{cite}
\usepackage{hyperref}
\hypersetup{
    pdfnewwindow=true,     
    colorlinks=true,       
    linkcolor={[rgb]{0,0,0}},          
    citecolor=black,        
    filecolor=black,     
    urlcolor=black      
}

\usepackage{mathtools} % SIMON
\usepackage{lmodern} % SIMON
\DeclareMathAlphabet{\mathpzc}{OT1}{pzc}{m}{it}
{\begin{list}{}%
         {\setlength{\leftmargin}{#1}}%
         \item[]%
}
{\end{list}}

\usepackage{pdfpages}

\graphicspath{{figures/}}

\begin{document}
\bstctlcite{IEEEexample:BSTcontrol}
\DeclarePairedDelimiter{\parentheses}{(}{)}

	\sloppy
\dashlinedash 2pt
\dashlinegap 2pt

\title{EEG-informed attended speaker extraction from recorded speech mixtures with application in neuro-steered hearing prostheses}

\author{Simon Van Eyndhoven, Tom Francart, and Alexander Bertrand,~\IEEEmembership{Member,~IEEE}.\\
\vspace{-0.7cm}
\thanks{This work was carried out at the ESAT Laboratory of KU Leuven, in the frame of KU Leuven Research Council BOF/STG-14-005, CoE  PFV/10/002 (OPTEC), Research Projects FWO nr. G.0931.14 `Design of distributed signal processing algorithms and scalable hardware platforms for energy-vs-performance adaptive wireless acoustic sensor networks', and \mbox{HANDiCAMS}. The project HANDiCAMS acknowledges the financial support of the Future and Emerging Technologies (FET) programme within the Seventh Framework Programme for Research of the European Commission, under FET-Open grant number: 323944. The scientific responsibility is assumed by its authors.
\par
S. Van Eyndhoven and A. Bertrand are with KU Leuven, Department of Electrical Engineering (ESAT), Stadius Center for Dynamical Systems, Signal Processing and Data Analytics, Kasteelpark Arenberg 10, box 2446, 3001 Leuven, Belgium (e-mail: \mbox{simon.vaneyndhoven@esat.kuleuven.be}, \mbox{alexander.bertrand@esat.kuleuven.be}). T. Francart is with KU Leuven, Department of Neurosciences, Research Group Experimental Oto-rhino-laryngology (e-mail: tom.francart@med.kuleuven.be).}}

\IEEEpubid{\begin{minipage}{\textwidth}\ \\[15pt]
	This paper is published in IEEE Transactions on Biomedical Engineering (2016) and is under copyright. Please cite this paper as:
	\newline S. Van Eyndhoven, T. Francart, and A. Bertrand, "EEG-informed attended speaker extraction from recorded speech mixtures with application in neuro-steered hearing prostheses", IEEE Transactions on Biomedical Engineering, vol. 64, no. 5, pp. 1045-1056, 2017.
\end{minipage}}

\maketitle
\begin{abstract}
\textit{Objective}: We aim to extract and denoise the attended speaker in a noisy, two-speaker acoustic scenario, relying on microphone array recordings from a binaural hearing aid, which are complemented with electroencephalography (EEG) recordings to infer the speaker of interest. \textit{Methods}: In this study, we propose a modular processing flow that first extracts the two speech envelopes from the microphone recordings, then selects the attended speech envelope based on the EEG, and finally uses this envelope to inform a multi-channel speech separation and denoising algorithm. \textit{Results}: Strong suppression of interfering (unattended) speech and background noise is achieved, while the attended speech is preserved. Furthermore, EEG-based auditory attention detection (AAD) is shown to be robust to the use of noisy speech signals. \textit{Conclusions}: Our results show that  AAD-based speaker extraction from microphone array recordings is feasible and robust, even in noisy acoustic environments, and without access to the clean speech signals to perform EEG-based AAD. \textit{Significance}: Current research on AAD always assumes the availability of the clean speech signals, which limits the applicability in real settings. We have extended this research to detect the attended speaker even when only microphone recordings with noisy speech mixtures are available. This is an enabling ingredient for new brain-computer interfaces and effective filtering schemes in neuro-steered hearing prostheses. Here, we provide a first proof of concept for EEG-informed attended speaker extraction and denoising.
\end{abstract}
\begin{IEEEkeywords}
EEG signal processing, speech enhancement, auditory attention detection, brain-computer interface, auditory prostheses, blind source separation, multi-channel Wiener filter
\end{IEEEkeywords}

\section{Introduction}

In order to guarantee speech intelligibility in a noisy, multi-talker environment, often referred to as a `cocktail party scenario', hearing prostheses can greatly benefit from effective noise reduction techniques \cite{dillon2001hearing,doclo2002gsvd}. While numerous and successful efforts have been made to achieve this goal, e.g. by incorporating the recorded signals of multiple microphones \cite{doclo2002gsvd,serizel2014low,doclo2009reduced}, many of these solutions strongly rely on the proper identification of the target speaker in terms of voice activity detection (VAD). In an acoustic scene with multiple competing speakers, this is a highly non-trivial task, complicating the overall problem of noise suppression. Even when a good speaker separation is possible, a fundamental problem that appears in such multi-speaker scenarios is the selection of the speaker of interest. To make a decision, heuristics have to be used, e.g., selecting the speaker with highest energy, or the speaker in the frontal direction. However, in many real-life scenarios, such heuristics fail to adequately select the attended speaker.

\IEEEpubidadjcol
Recently however, auditory attention detection (AAD) has become a popular topic in neuroscientific and audiological research. Different experiments have confirmed the feasibility of a decoding paradigm that, based on recordings of brain activity such as the electroencephalogram (EEG), detects to which speaker a subject attends in an acoustic scene with multiple competing speech sources \cite{o2014attentional,ding2012emergence,golumbic2013mechanisms,mesgarani2012selective,biesmansauditory,mirkovic2015decoding}. A major drawback of all these experiments is that they place strict constraints on the methodological design, which limits the practical use of their results. More precisely all of the proposed paradigms employ the separate `clean' speech sources that are presented to the subjects (to correlate their envelopes to the EEG data), a condition which is never met in realistic acoustic applications such as hearing prostheses, where only the speech mixtures as observed by the device's local microphone(s) are available. In \cite{aroudiaad} it is reported that the detection performance drops substantially under the effect of crosstalk or uncorrelated additive noise on the reference speech sources that are used for the auditory attention decoding. It is hence worthwhile to further investigate AAD that is based on \textit{mixtures} of the speakers, such as in the signals recorded by the microphones of a hearing prosthesis.

Nonetheless, devices such as neuro-steered hearing prostheses or other brain-computer interfaces (BCIs) that implement AAD, can only be widely applied in realistic scenarios if they can operate reliably in these noisy conditions. End users with (partial) hearing impairment could greatly benefit from neuro-steered speech enhancement and denoising technologies, especially if they are implemented in compact mobile devices. EEG is the preferred choice for these emerging solutions, due to its cheap and non-invasive nature \cite{mihajlovic2015wearable,looney2012ear,bleichner2015exploring,norton2015soft,bertrand2014distributed,casson2010wearable}. Many research efforts have been focused on different aspects of this modality to enable the development of small scale, wearable EEG devices. Several studies have addressed the problem of wearability and miniaturization \cite{looney2012ear,bleichner2015exploring,norton2015soft,bertrand2014distributed}, data compression and power consumption \cite{bertrand2014distributed,casson2010wearable}.

In this study, we combine EEG-based auditory attention detection and acoustic noise reduction, to suppress interfering sources (including the unattended speaker) from noisy multi-microphone recordings in an acoustic scenario with two simultaneously active speakers. Our algorithm enhances the attended speaker, using EEG-informed AAD, based \textit{only} on the microphone recordings of a hearing prosthesis, i.e., without the need for the clean speech signals\footnote{We still use clean speech signals to design the EEG decoder in an initial training or calibration phase. However, once this decoder is obtained, our algorithm operates directly on the microphone recordings, without using the original clean speech signals as side-channel information.}. The final goal is to have a computationally cheap processing chain that takes microphone and EEG recordings from a noisy, multi-speaker environment at its input and transforms these into a denoised audio signal in which the attended speaker is enhanced, and the unattended speaker is suppressed. To this end, we reuse experimental data from the AAD experiment in \cite{biesmansauditory} and use the same speech data as in \cite{biesmansauditory} to synthesize microphone recordings of a binaural hearing aid, based on publicly available head-related transfer functions which were measured with real hearing aids \cite{kayser2009database}. 
As we will show further on, non-negative blind source separation is a convenient tool in our approach, as we need to extract the speech envelopes from the recorded mixtures. To this end, we rely on \cite{MNICAconf}, where a low-complexity source separation algorithm is proposed that can operate at a sampling rate that is much smaller than that of the microphone signals, which is very attractive from a computational point of view. We investigate the robustness of our processing scheme by adding varying amounts of acoustic interference and testing different speaker setups. 

The outline of the paper is as follows. In section \ref{sec:probstatement}, we give a global overview of the problem and an introduction to the different aspects we will address; in section \ref{sec:algorithm} we explain the techniques for non-negative blind source separation, and cover the extraction of the attended speech from (noisy) microphone recordings; in section \ref{sec:experiment} we describe the conducted experiment; in section \ref{sec:results} we elaborate on the results of our study; in section \ref{sec:discussion} we discuss these results and consider future research directions; in section \ref{sec:conclusion} we conclude the paper.

\section{Problem statement}\label{sec:probstatement}
\subsection{Noise reduction problem}\label{subsec:NRprob}
We consider a (binaural) hearing prosthesis equipped with multiple microphones, where the signal observed by the $i$-th microphone is modeled as a convolutive mixture: 
\begin{align}
\label{eq:convolmix}
	m_{i}[t] &= (h_{i1}\ast s_{1})[t] + (h_{i2}\ast s_{2})[t] + v_i[t]\\
	&= x_{i1}[t] + x_{i2}[t] + v_i[t].
\end{align}

\noindent In (\ref{eq:convolmix}), $m_{i}[t]$ denotes the recorded signal at microphone $i$, which is a superposition of contributions $x_{i1}[t]$ and $x_{i2}[t]$ of both speech sources and a noise term $v_i[t]$. $x_{i1}[t]$ and $x_{i2}[t]$ are the result of the convolution of the clean (`dry') speech signals $s_{1}[t]$ and $s_{2}[t]$ with the head-related impulse responses (HRIRs) $h_{i1}[t]$ and $h_{i2}[t]$, respectively. These HRIRs are assumed to be unknown and model the acoustic propagation path between the source and the $i$-th microphone, including head-related filtering effects and reverberation. The term $v_i[t]$ bundles all background noise impinging on microphone $i$ and contaminating the recorded signal. 

Converting (\ref{eq:convolmix}) to the (discrete) frequency domain, we get
\begin{align}
\label{eq:convolmix_freq}
	M_{i}(\omega_j) &= H_{i1}(\omega_j) S_{1}(\omega_j) + H_{i2}(\omega_j) S_{2}(\omega_j) + V_i(\omega_j)\\
	&= X_{i1}(\omega_j) + X_{i2}(\omega_j) + V_i(\omega_j)
\end{align}
for all frequency bins $\omega_j$. In (\ref{eq:convolmix_freq}), $M_{i}(\omega_j)$, $S_{1}(\omega_j)$, $S_{2}(\omega_j)$ and $V_i(\omega_j)$ are representations of the recorded signal at microphone $i$, the two speech sources and the noise at frequency $\omega_j$, respectively. $H_{i1}(\omega_j)$ and $H_{i2}(\omega_j)$ are the frequency-domain representations of the HRIRs, which are often denoted as head-related transfer functions (HRTFs). All microphone signals and speech contributions can then be stacked in vectors $\mathbf{M}(\omega_j) = \left[M_{1}(\omega_j) \dots M_{K}(\omega_j)\right]^T$, $\mathbf{X}_1(\omega_j) = \left[X_{11}(\omega_j) \dots X_{K1}(\omega_j)\right]^T$ and $\mathbf{X}_2(\omega_j) = \left[X_{12}(\omega_j) \dots X_{K2}(\omega_j)\right]^T$, where $K$ is the number of available microphones. Our aim is to enhance the attended speech component and suppress the interfering speech and noise in the microphone signals. More precisely, we arbitrarily select a reference microphone (e.g. $r=1$) and, assuming without loss of generality that $s_{1}[t]$ is the attended speech, try to estimate $X_{r1}(\omega_j)$ by filtering $\mathbf{M}(\omega_j)$, which is the full set of microphone signals\footnote{In the case of a binaural hearing prosthesis, we assume that the microphone signals recorded at the left and right ear can be exchanged between both devices, e.g., over a wireless link \cite{doclo2009reduced}.}. Hereto, a linear minimum mean-squared error (MMSE) cost criterion is used \cite{doclo2002gsvd,serizel2014low}:
\begin{equation}\label{eq:costfunction}
J(\mathbf{W}(\omega_j)) = \operatorname{E}\left\{\lvert \mathbf{W}(\omega_j)^H\mathbf{M}(\omega_j) - X_{r1}(\omega_j) \rvert^2\right\}
\end{equation}
in which $\mathbf{W}$ is a $K$-channel filter, represented by a $K$-dimensional complex-valued vector, where the superscript $H$ denotes to the conjugate transpose. Note that a different $\mathbf{W}$ is selected for each frequency bin, resulting in a spatio-spectral filtering, which is equivalent to a convolutive spatio-temporal filtering when translated to the time-domain. In section \ref{subsec:MWF}, we will minimize (\ref{eq:costfunction}) by means of the so-called multi-channel Wiener filter (MWF). 

Up to now, it is not known which of the speakers is the target or attended speaker. To determine this, we need to perform auditory attention detection (AAD), as described in the next subsection. Furthermore, the MWF paradigm requires knowledge of the times at which this attended speaker is active. To this end, we need a speaker-dependent voice activity detection (VAD), which will be discussed in subsection \ref{subsec:VAD}. We only have access to the envelopes of the microphone signals, which contain significant crosstalk due to the presence of two speakers. Hence, relying on these envelopes would lead to suboptimal performance (i.e. misdetections of the VAD), motivating the use of an intermediate step to obtain better estimates of these envelopes.  As stated, we employ non-negative blind source separation to obtain more accurate estimates of the envelopes, which will prove to relax the VAD problem (see \ref{subsec:MNICA}). 

\subsection{Auditory attention detection (AAD) problem}\label{subsec:AADprob}
In (\ref{eq:convolmix}), either $s_{1}[t]$ or $s_{2}[t]$ can be the attended speech. Earlier studies showed that the low frequency variations of speech envelopes (between approximately 1 and 9 Hz) are encoded in the evoked brain activity \cite{aiken2008human,pasley2012reconstructing}, and that this mapping differs whether the speech is attended to by the subject (or not) in a multi-speaker environment \cite{ding2012emergence,golumbic2013mechanisms,mesgarani2012selective,ding2012neural,kerlin2010attentional}. This mapping can be reversed to categorize the attention of a listener from recorded brain activity. In brief, the AAD paradigm works by first training a spatiotemporal filter (decoder) on the recorded EEG data to reconstruct the envelope of the attended speech by means of a linear regression \cite{o2014attentional,biesmansauditory,mirkovic2015decoding,aroudiaad}. This decoder will reconstruct an auditory envelope, by integrating the measured brain activity across $\kappa$ channels and for $\tau_{max}$ different lags, described by
\begin{align}
\label{eq:reconstruction_generic1}
\widehat{s}_{{}_A}[n] &= \sum_{\tau = 0}^{\tau_{max}}\sum_{k = 1}^{\kappa}r_k[n+\tau]d_k[\tau]
\end{align}
\noindent in which $r_k[n]$ is the recorded EEG signal at channel $k$ and time $n$, $d_k[\tau]$ is the decoder weight for channel $k$ at a post-stimulus lag of $\tau$ samples, and $\hat{s}_{{}_A}[n]$ is the reconstructed attended envelope at time $n$. We can rewrite this expression in matrix notation, as $\mathbf{\widehat{s}}_{{}_A} = \mathbf{R\,d}$, in which $\mathbf{\widehat{s}}_{{}_A}$ is a vector containing the samples of the reconstructed envelope, $\mathbf{d} = [\,d_0[0] \dots d_0[\tau_{max}]\dots d_{\kappa}[0] \dots d_{\kappa}[\tau_{max}]\,]^T$ is a vector with the stacked spatiotemporal weights, of length channels $\times$ lags, and where the matrix with EEG measurements is structured as $\mathbf{R} = [\, \mathbf{r}_1\dots \mathbf{r}_N \,]^T$, where there is a vector $\mathbf{r}_n = [\,r_0[n]\dots r_0[n+\tau_{max}]\dots r_{\kappa}[n]\dots r_{\kappa}[n+\tau_{max}]]^T$ for every sample $n = 1\dots N$ of the envelope. We find the decoder by solving the following optimization problem: 
\begin{align}
	\mathbf{\widehat{d}} &= \text{arg}\;\underset{\mathbf{d}}{\text{min}}\: \lVert\mathbf{\,\widehat{s}}_{{}_A}-\mathbf{s}_{{}_A}\rVert^2\\
	&= \text{arg}\;\underset{\mathbf{d}}{\text{min}}\: \lVert \mathbf{R\, d}-\mathbf{s}_{{}_A}\rVert^2
	\label{eq:reconstruction_LS1}
\intertext{in which $\mathbf{s}_{{}_A}$ is the real envelope of the attended speech. Using classical least squares, we compute the decoder weights as}
	\mathbf{\widehat{d}} &= (\mathbf{R}^{T}\!\mathbf{R})^{-1}\mathbf{R}^{T}\!\mathbf{s}_{{}_A}\label{eq:reconstruction_LS2}.
\end{align}
The matrix $\mathbf{R}^{T}\!\mathbf{R}$ represents the sample autocorrelation matrix of the EEG data (for all channels and considered lags) and $\mathbf{R}^{T}\!\mathbf{s}_{{}_A}$ is the sample cross-correlation of the EEG data and the attended speech envelope. Hence, the decoder $\mathbf{\widehat{d}}$ is trained to optimally reconstruct the envelope $\mathbf{s}_{{}_A}$ of the attended speech sources. If the sample correlation matrices are estimated on too few samples, a regularization term can be used, like in \cite{mirkovic2015decoding}. As motivated in subsection \ref{subsec:AADexp}, we omitted regularization in this study.

The decoding is successful if the decoder reconstructs an envelope that is more correlated with the envelope of the attended speech than with that of the unattended speech. Mathematically, this translates to $r_{{}_{A}}>r_{{}_{U}}$, in which $r_{{}_{A}}$ and $r_{{}_{U}}$ are the Pearson correlation coefficients of the reconstructed envelope $\mathbf{\widehat{s}}_{{}_A}$ with the envelopes of the attended and unattended speech, respectively. In this paper, rather than requiring the separate speech envelopes to be available, we make the assumption that we only have access to the recorded microphone signals (except for the training of the EEG decoder based on (\ref{eq:reconstruction_LS2})). In section \ref{sec:algorithm}, we address the problem of speech envelope extraction from the speech mixtures in the microphone signals, to still be able to perform AAD using the approach explained above.

\section{Algorithm pipeline}\label{sec:algorithm}
\begin{figure}
	\centering
	\includegraphics[width=.5\textwidth]{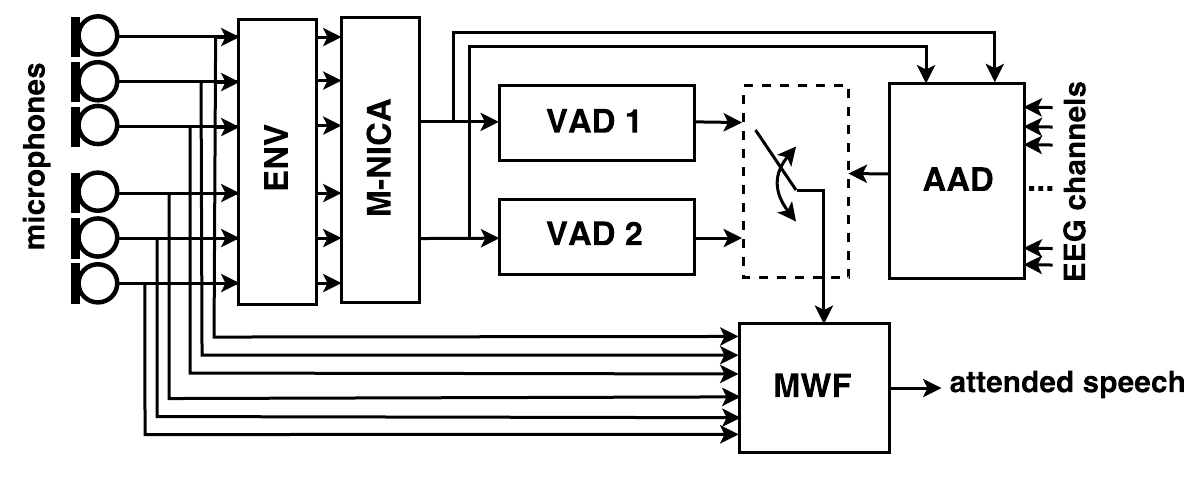}
	\caption{Pipeline of the proposed processing flow.}
	\label{fig:blockdiagram}
\end{figure}

Here, we propose a modular processing flow that comprises a number of steps towards the extraction and denoising of the attended speech, shown as a block diagram in \figurename~\ref{fig:blockdiagram}. We compute the energy envelopes of the recorded microphone mixtures (represented by the `\textsc{env}'-block and explained in subsection \ref{subsec:energyconversion}) and use the multiplicative non-negative independent component analysis (M-NICA) algorithm to estimate the original speech envelopes from these mixtures (subsection \ref{subsec:MNICA}). These speech envelopes are fed into the AAD processing block described in previous subsection, which will indicate one of both as belonging to the attended speaker, based on the EEG recording (arrows on the right). Voice activity detection is carried out on the estimated envelopes, and the VAD track that is selected during AAD serves as input to the multi-channel Wiener filter (subsection \ref{subsec:VAD}). The MWF filters the set of microphone mixtures, based on this VAD track, yielding one enhanced speech signal at the output (subsection \ref{subsec:MWF}).

\subsection{Conversion to energy domain (ENV)}\label{subsec:energyconversion}
In order to apply the AAD algorithm described in subsection \ref{subsec:AADprob}, we need the envelopes of the individual speech sources. Since we are only interested in the speech envelopes, we will work in the energy domain, allowing to solve a source separation problem at a much lower sampling rate than the original sampling rate of the microphone signals. Furthermore, energy signals are non-negative, which can be exploited to perform real-time source separation based only on second-order statistics \cite{bertrand2010blind}, rather than higher-order statistics as in many of the standard independent component analysis techniques. These two ingredients result in a computationally efficient algorithm, which is important when it is to be operated in a battery-powered miniature device such as a hearing prosthesis. A straightforward way to calculate an energy envelope is by squaring and low-pass filtering a microphone signal, i.e., for microphone $i$ this yields the energy signal
\begin{equation}
\label{eq:shorttimeenergy}
E_{m_i}[n] = \frac{1}{T}\sum\limits_{w=1}^{T} \! m_i[n\,T+w]^2
\end{equation}
in which $n$ is the sample index of the energy signal, $T$ is the number of samples (window length) to compute the short-time average energy $E_{m_i}[n]$, which estimates the real microphone energy, $\operatorname{E}\{m^2_i[n\,T]\}$.

Based on (\ref{eq:convolmix}), and assuming the source signals are independent, we can model the relationship between the envelopes of the speech sources and the microphone signals as an approximately linear, instantaneous mixture of energy signals:
\begin{equation}
\label{eq:linearmixing}
\mathbf{E}_m[n] \approx \mathbf{A}\,\mathbf{E}_s[n] + \mathbf{E}_v[n]\,.
\end{equation}
Here, the short-time energies of the $K$ microphone signals and the $S$ speech sources are stacked in the time-varying vectors $\mathbf{E}_m[n]$ and $\mathbf{E}_s[n]$, respectively, and are related through the $K\times S$ mixing matrix $\mathbf{A}$, defining the overall energy attenuation between every speech source and every microphone. Similarly, the short-term energies of the $N$ noise components that contaminate the microphone signals are represented by the vector $\mathbf{E}_v[n]$. For infinitely large $T$ and infinitely narrow impulse responses, (\ref{eq:linearmixing}) is easily shown to be exact. For HRIRs of a finite duration and for finite $T$, it is a quite rough approximation, but we found that it still provides a useful basis for the subsequent algorithm that aims to estimate the original speech envelopes from the mixtures, as we succeed to extract the original speech envelopes reasonably well (see next subsection and section \ref{sec:results}).
The literature also reports experiments where the approximation in (\ref{eq:linearmixing}) has succesfully been used as a mixing model for separation of speech envelopes, even in reverberant environments with longer impulse responses than the HRIRs that are used here \mbox{\cite{MNICAconf,chouvardas2015distributed}}. 

\subsection{Speech envelope extraction from mixtures (M-NICA)}\label{subsec:MNICA}
The M-NICA algorithm is a technique that exploits the non-negativity of the underlying sources \cite{bertrand2010blind} to solve blind source separation (BSS) problems in an efficient way. It demixes a set of observed signals, that is the result of a linear mixing process, into its separate, nonnegative sources. Under the assumption that the source signals are independent, non-negative, and well-grounded\footnote{A signal is well-grounded if it attains zero-valued samples with finite probability \cite{bertrand2010blind}.}, it can be shown that a perfect demixing is obtained by a demixing matrix that decorrelates the signals while preserving non-negativity. Similar to \cite{MNICAconf}, we will employ the M-NICA algorithm, to find an estimate of $\mathbf{E}_s[n]$ from $\mathbf{E}_m[n]$ in (\ref{eq:linearmixing}). The algorithm consists of an iterative interleaved application of a multiplicative decorrelation step (preserving the non-negativity), and a subspace projection step (to re-fit the data to the model). An in-depth description of the M-NICA algorithm is available in \cite{bertrand2010blind}, which also includes a sliding-window implementation for real-time processing. Attractive properties of M-NICA are that it relies only on 2\textsuperscript{nd} order statistics (due to the non-negativity constraints) and that it operates at the low sampling rate of the envelopes. These features foster the use of M-NICA, as the algorithm seems to be well matched to the constraints of the target application, namely the scarce computational resources and the required real-time operation. Note that the number of speech sources must be known a priori. In practice, we could estimate this number by a singular value decomposition \cite{MNICAconf}. We will refer to $\mathbf{E}_m[n]$ and $\mathbf{\widehat{E}}_s[n]$ as the microphone envelopes and demixed envelopes, respectively, where ideally $\mathbf{\widehat{E}}_s[n] = \mathbf{E}_s[n]$. As with most BSS techniques, a scaling and permutation ambiguity remains, i.e., the ordering of the sources and their energy cannot be found, since they can be arbitrarily changed if a compensating change is made in the mixing matrix. In real-time, adaptive applications, these ambiguities stay more or less the same as time progresses and are of little importance (see \cite{MNICAconf}, where an adaptive implementation of M-NICA is tested on speech mixtures). It is noted that, to perform M-NICA on (\ref{eq:linearmixing}), the matrix $\mathbf{A}$ should be well-conditioned in the sense that it should have at least two singular values that are significantly larger than 0. This means that the energy contribution of each speech source should be differently distributed over the $K$ microphones. In \cite{MNICAconf} and \cite{chouvardas2015distributed}, this was obtained by placing the microphone several meters apart, which is not possible in our application of hearing prostheses. However, we use microphones that are on both sides of the head, such that the head itself acts as an angle-dependent attenuator for each speaker location. This results in a different spatial energy pattern for each speech source and hence in a well-conditioned energy mixing matrix $\mathbf{A}$.

\subsection{Multi-channel Wiener filter (MWF)}\label{subsec:MWF}
For the sake of conciseness, we will omit the frequency variable $\omega_j$ in the remainder of the text. The solution that minimizes the cost function in (\ref{eq:costfunction}) is the multi-channel Wiener filter $\widehat{\mathbf{W}}$ \cite{doclo2002gsvd,serizel2014low,doclo2009reduced}, found as
\begin{align}\label{eq:mmsesolution}
\widehat{\mathbf{W}} &= \text{arg}\;\underset{\mathbf{W}}{\text{min}}\operatorname{E}\left\{\lvert \mathbf{W}^H\mathbf{M} - X_{r1} \rvert^2\right\}\\ 
&= \mathbf{R}_{mm}^{-1}\mathbf{R}_{xx}\,\mathbf{e}_r\\
&= (\mathbf{R}_{xx}+\mathbf{R}_{vv})^{-1}\mathbf{R}_{xx}\,\label{eq:mmsesolution3}\mathbf{e}_r
\end{align}
in which $\mathbf{R}_{mm}$ is the $K\times K$ autocorrelation matrix $\operatorname{E}\{\mathbf{MM}^H\}$ of the microphone signals and $\mathbf{R}_{xx}$ is the $K\times K$ speech autocorrelation matrix $\operatorname{E}\{\mathbf{X}_1\mathbf{X}_1^H\}$, where the subscript 1 refers to the attended speech. Likewise, $\mathbf{R}_{vv}$ is the $K\times K$ autocorrelation matrix of the undesired signal component. Note that the MWF will estimate the speech signal $S_{1}$ as it is observed by the selected reference microphone, i.e., it will estimate $H_{r1} S_{1}$, assuming the r-th microphone is selected as the reference. Hence, $\mathbf{e}_r$ is the $r$-th column of an identity matrix, which selects the $r$-th column of $\mathbf{R}_{xx}$ corresponding to this reference microphone.

The matrix $\mathbf{R}_{xx}$ is unknown, but can be estimated as $\mathbf{R}_{xx} = \mathbf{R}_{mm} - \mathbf{R}_{vv}$, with $\mathbf{R}_{mm}$ the `speech plus interference' autocorrelation matrix, equal to $\operatorname{E}\{\mathbf{MM}^H\}$ when measuring during periods in which the attended speaker is active. Likewise, $\mathbf{R}_{vv}$ can be found as $\operatorname{E}\{\mathbf{MM}^H\}$, during periods when the attended speaker is silent. All of the mentioned autocorrelation matrices can be estimated by means of temporal averaging in the short-time Fourier transform domain.
Note that more robust ways exist to estimate $\mathbf{R}_{xx}$, compared to the straightforward subtraction described here. The MWF implementation we employed uses a generalized eigenvalue decomposition (GEVD) to find a rank-1 approximation of $\mathbf{R}_{xx}$ as in \cite{serizel2014low}. The rationale behind this is that the MWF aims to enhance a single speech source (corresponding to the attended speaker) while suppressing all other acoustic sources (other speech and noise). Since $\mathbf{R}_{xx}$ only captures a single speech source, it should have \mbox{rank 1}. 

Applying the MWF corresponds to computing (\ref{eq:mmsesolution3}) and performing the filtering $\mathbf{W}^H\mathbf{M}$ for each frequency $\omega_j$ and each time-window in the short-time Fourier domain. Finally, the resulting output in the short-time Fourier domain can be transformed back to the time domain again. In practice, this is often done using a weighted overlap-add (WOLA) procedure \cite{BertrandIWAENC10}.

As mentioned above, when estimating $\mathbf{R}_{xx}$ and $\mathbf{R}_{nn}$ from the microphone signals $\mathbf{M}$, we rely on a good identification of periods or frames in which both (attended) speech and interference are present (to estimate the speech-plus-interference autocorrelation $\mathbf{R}_{mm}$)  versus periods during which only interference is recorded (to estimate the interference-only correlation $\mathbf{R}_{vv}$). Making this distinction corresponds to voice activity detection, which we discuss next.

\subsection{Voice activity detection (VAD)}\label{subsec:VAD}
The short-time energy of a speech signal gives an indication at what times the target speech source is (in)active.
A simple voice activity detection (VAD) algorithm consists of thresholding the energy envelope of the target speech signal. Note that in our target application, the speech envelopes are also used for AAD. After applying M-NICA on the microphone envelopes, we find two demixed envelopes, which serve as better estimates of the real speech envelopes. Based on the correlation with the reconstructed envelope $\mathbf{\widehat{s}_{{}_A}}$ from the AAD decoder in (\ref{eq:reconstruction_generic1}), one of these demixed envelopes will be identified as the envelope of the attended speech source. This correlation can be computed efficiently in a recursive sliding-window fashion, to update the AAD decision over time, which is represented by a time-varying switch in \figurename~\ref{fig:blockdiagram}. For each AAD decision, the chosen envelope segment is then thresholded sample-wise for voice activity detection. Ideally, the envelope segments on which the VAD is applied all originate from the attended envelope, although sometimes the unattended envelope may be wrongfully selected, depending on the AAD decisions that are made. This will lead to VAD errors, which will have an impact on the denoising and speaker extraction performance of the MWF.
\section{Experiment}\label{sec:experiment}
For every pair of speech sources (1 attended and 1 unattended), we performed the following steps: \begin{enumerate}
	\item compute the microphone signals, according to (\ref{eq:convolmix})
	\item find the energy-envelope of the microphone signals, as described in subsection \ref{subsec:energyconversion}
	\item demix the microphone envelopes with M-NICA, as described in subsection \ref{subsec:MNICA}
	\item find the VAD track for the attended speech source, as described in subsection \ref{subsec:VAD}, based on the results of the auditory attention task described in \ref{subsec:AADexp}
	\item compute the MWF for the attended speech source, as described in subsection \ref{subsec:MWF}, based on the AAD-selected VAD track from step 4
	\item filter the microphone signals with this MWF using a WOLA procedure, to enhance the attended speech source
\end{enumerate}
Furthermore, we also investigate the overall performance if step 3 is skipped, i.e., if we use the plain microphone envelopes without demixing them with M-NICA. In that case, we manually pick the two microphone envelopes that are already most correlated to either of both speakers. Note that this is a best-case scenario that cannot be implemented in practice.

\subsection{Microphone recordings}\label{subsec:simulation}
\noindent We synthesized the microphone array recordings using a public database of HRIRs that were measured using six behind-the-ear microphones (three microphones per ear) \cite{kayser2009database}. Each HRIR represents the microphone impulse responses for a source at a certain azimuthal angle relative to the head orientation and at 3 meters distance from the microphone. The HRIRs were recorded in an anechoic room and had a length of 4800 samples at 48 kHz. As speech sources, we used Dutch narrated stories (each with a length of approximately six minutes and a sampling rate of 44.1 kHz), that previously served as the auditory stimuli in the AAD-experiment in \cite{biesmansauditory}. 

To determine the robustness of our scheme, we included noise in the acoustic setup. We synthesize the microphone signals for several speaker positions, ranging from -90$^{\circ}$ to 90$^{\circ}$. The background noise is formed by adding five uncorrelated multi-talker noise sources $n_{k}[n]$ at positions $-90\,^{\circ}$, $-45\,^{\circ}$, 0$\,^{\circ}$, 45$\,^{\circ}$ and 90$\,^{\circ}$ and at 3 meters distance, each with a long-term power $P_{N_k} = 0.1P_{s}$, in which $P_{s}$ is the long-term power of a single speech source. Note that these noise sources were not present in the stimuli used in the AAD experiment, and are only added here to illustrate the robustness of M-NICA to a possible noise term in (\ref{eq:linearmixing}), and to illustrate the denoising capabilities of the MWF. We convolve the two speech signals and five noise signals with the corresponding HRIRs to synthesize the microphone signals described in (\ref{eq:convolmix}). The term $v_i[n]$ thus represents all noise contributions and is calculated as $\sum_{k}^{} (h_{ik}\ast n_{k})[n]$, where the five $h_{ik}[n]$ are the HRIRs for the noise sources.

In our study, we evaluate the performance for 12 representative setups with varying spatial angle between the two speaker locations. Taking $0\,^{\circ}$ as the direction in front of the subject wearing the binaural hearing aids, the angular position pairs of the speakers are  $-90\,^{\circ}$ and $90\,^{\circ}$, $-75\,^{\circ}$ and $75\,^{\circ}$, $-90\,^{\circ}$ and $30\,^{\circ}$, $-60\,^{\circ}$ and $60\,^{\circ}$, $-90\,^{\circ}$ and $0\,^{\circ}$, $-45\,^{\circ}$ and $45\,^{\circ}$, $-90\,^{\circ}$ and $-30\,^{\circ}$, $-60\,^{\circ}$ and $0\,^{\circ}$, $-30\,^{\circ}$ and $30\,^{\circ}$, $-90\,^{\circ}$ and $-60\,^{\circ}$, $-60\,^{\circ}$ and $-30\,^{\circ}$, and $-15\,^{\circ}$ and $15\,^{\circ}$.

\vspace{-3pt}

\subsection{AAD experiment}\label{subsec:AADexp}
\noindent The EEG data originated from a previous study \cite{biesmansauditory}, in which 16 normal hearing subjects participated in an audiologic experiment to investigate auditory attention detection. In every trial, a pair of competing speech stimuli (1 out of 4 pairs of narrated Dutch stories, at a sampling rate of 8 kHz) is simultaneously presented to the subject to create a cocktail party scenario; the cognitive task requires the subject to attend to one story for the complete duration of every trial. We consider a subset of the experiment in \cite{biesmansauditory}, in which the presented speech stimuli have a contribution to each ear - after filtering them with in-the-ear HRIRs for sources at \mbox{-90$^{\circ}$} and 90$^{\circ}$ - in order to obtain a dataset of EEG-responses that is more representative for realistic scenarios. That is, both ears are presented with a (different) mixture of both speakers, mimicking the acoustic filtering by the head as if the speakers were located left and right of the subject. For every trial, the recorded EEG is then sliced in frames of 30 seconds, followed by the training of the AAD decoder and detection of the attention for every frame, in a leave-one-frame-out cross-validation fashion. We use the approach of \cite{biesmansauditory}, where a single decoder is estimated by computing (\ref{eq:reconstruction_LS2}) once over the full set of training frames, i.e., a single $\mathbf{R}^{T}\!\mathbf{R}$ and $\mathbf{R}^{T}\!\mathbf{s}_{{}_A}$ matrix is calculated over all samples in the training set. This is opposed to the method in \cite{o2014attentional}, where a decoder is estimated for each training frame separately, and the averaged decoder is then applied to the test frame. In \cite{biesmansauditory}, it was demonstrated that this approach is sensitive to a manually tuned regularization parameter and may affect performance, which is why we opted for the former method.
The performance of the decoders depends on the method of calculating the envelope $\mathbf{s_{{}_A}}$ of the attended speech stimulus. In \cite{biesmansauditory}, it was found that amplitude envelopes lead to better results than energy envelopes. For the present study, we work with energy envelopes (as described in subsection \ref{subsec:energyconversion}) and take the square root to convert to amplitude envelopes, when computing the correlation coefficients in the AAD task.

The present study inherits the recorded EEG data from the experiment described above, and assumes that decoders can be found during a supervised training phase in which the clean speech stimuli are known\footnote{Note that in a real device, only one final decoder would need to be available (obtained after a training phase).}. Throughout our experiment, we train the decoders per individual subject on the EEG data and the corresponding envelope segments of the attended speech stimuli, calculated by taking the absolute value of the original speech signals and filtering between 1 and 9.5 Hz (equiripple finite impulse response filter, -3 dB at 0.5 and 10 Hz). Contrary to \cite{o2014attentional}, attention during the trials was balanced over both ears, so that no ear-specific biasing could occur during training of the decoder. 

The trained decoder can then be used to detect to which speaker a subject attends, as explained in subsection \ref{subsec:AADprob}. We perform the auditory attention detection procedure with the same recorded EEG data (using leave-one-frame-out cross-validation) which is fed through the pre-trained decoder, and then correlated with different envelopes to eventually perform the detection over frames of 30 seconds. In order to assess the contribution of the M-NICA algorithm to the overall performance, we consider two options: either the two demixed envelopes or the two microphone envelopes that have the highest correlation with either of the speech sources' envelopes are correlated to the EEG decoder's output $\mathbf{\widehat{s}}_{{}_A}$. The motivation for the latter option is that in some microphones, one of both speech sources will be prevalent, and we can take the envelope of such a microphone signal as a (poor) estimate of the envelope of that speech source. This will lead to the best-case performance that can be expected with the use of envelopes of the microphones, without using an envelope demixing algorithm. 

\vspace{-6pt}

\subsection{Preprocessing and parameter selection}\label{subsec:params}
Speech fragments are normalized over the full length to have equal energy. All speech sources and HRIRs were resampled to 16 kHz, after which we convolved them pairwise and added the resulting signals to find the set of microphone signals. 

The window length $T$ in (\ref{eq:shorttimeenergy}) is chosen so that the energy envelopes are sampled at 20 Hz. To find the \mbox{short-term} amplitude in a certain bandwidth, we take the square root of all energy-like envelopes and filter them between 1 and 9.5 Hz before employing them to decode attention in the EEG epochs. Likewise, all $\kappa = $ 64 EEG channels are filtered in this frequency range and downsampled to 20 Hz. As in \cite{o2014attentional}, $\tau_{max}$ in (\ref{eq:reconstruction_generic1}) is chosen so that it corresponds to 250 ms poststimulus. For a detailed overview of the data acquisition and EEG decoder training, we refer to \cite{biesmansauditory}.

VAD tracks for the envelopes of both the attended and unattended speech are binary triggers (`on' or `off'), that are 1 when the energy envelope surpasses the chosen threshold. The value for this threshold was determined as the one that would lead to the highest median SNR at the MWF output, for a virtual subject with an AAD accuracy of 100\% and in the absence of noise sources. After exhaustive testing, this value was set to $0.05\,\text{max}\left\{\mathbf{\widehat{E}}_s\right\}$ and $0.10\,\text{max}\left\{\mathbf{E}_m\right\}$ for the demixed and microphone envelopes, respectively (see subsection \ref{subsec:results-denoising}).
We form one hybrid VAD track by selecting and concatenating segments of 30 seconds of these two initial tracks, according to the AAD decision that was made in the same 30-second trial of the experiment, as described in subsection \ref{subsec:AADexp}. This corresponds to a non-overlapping sliding window implementation with a window length of 30 seconds (note that the AAD decision rate can be increased by using an overlapping sliding window with a window shift that is smaller than the window length). Thus, this overall VAD track, which is an input to the MWF, follows the switching behavior of the AAD-driven module shown in \figurename~\ref{fig:blockdiagram}. 

The MWF is applied on the binaural set of six microphone signals (resampled to 8 kHz, conform to the presented stimuli in the EEG experiment), through WOLA filtering with a square-root Hann window and FFT-length of 512. Likewise, the VAD track is expanded to match this new sample frequency. 

For this initial proof of concept, both M-NICA and the MWF are applied in batch mode on the signals, meaning that the second-order signal statistics are measured over the full signal length. In practice, an adaptive implementation will be necessary, which is beyond the scope of this paper. However, performance of M-NICA and MWF under adaptive sliding-window implementations have been reported in \cite{bertrand2010blind,BertrandIWAENC10}, where a significant - but acceptable - performance decrease is observed due to the estimation of the second-order statistics over finite windows. Therefore, the reported results in this paper should be interpreted as upper limits for the achievable performance with an adaptive system. For envelope demixing, 100 iterations of M-NICA are used.
\section{Results}\label{sec:results}

\subsection{Performance measures}
The microphone envelopes at the algorithm's input have considerable contributions of both speech sources. What is desired - as well for the VAD block as for the AAD block - is a set of demixed envelopes that are well-separated in the sense that each of them only tracks the energy of a single speech source, and thus has a high correlation with only one of the clean speech envelopes, and a low residual correlation with the other clean speech envelope. Hence, we adopt the following measure: $\Delta r_{{}_{HL}}$ is the difference $r_{{}_{H}} - r_{{}_{L}}$ between the highest Pearson correlation that exists between a demixed or microphone envelope and a speech envelope and the lowest Pearson correlation that is found between any other envelope and this speech envelope. E.g. for speech envelope 1, if the envelope of microphone 3 has the highest correlation with this speech envelope, and the envelope of microphone 5 has the lowest correlation, we assign these correlations to $r_{{}_{H}}$ and $r_{{}_{L}}$, respectively. For every angular separation of the two speakers, we will consider the average of $\Delta r_{{}_{HL}}$ over all speech fragments of all source combinations, and over all tested speaker setups that correspond to the same separation (see subsection \ref{subsec:simulation}). An increase of this parameter indicates a proper behavior of the M-NICA algorithm, i.e., it measures the degree to which the microphone envelopes (`a priori' $\Delta r_{{}_{HL}}$) or demixed envelopes (`a posteriori' $\Delta r_{{}_{HL}}$) are separated into the original speech envelopes. Note that for the `a priori' value, we select the microphones which already have the highest $\Delta r_{{}_{HL}}$ in order to provide a fair comparison. In practice, it is not known which microphone yields the highest $\Delta r_{{}_{HL}}$'s, which is another advantage of M-NICA: it provides only two signals in which this measure already maximized.

The decoding accuracy of the AAD algorithm is the percentage of trials that are correctly decoded. Analogous to the criterion in subsection \ref{subsec:AADprob}, if the reconstructed envelope $\mathbf{\widehat{s}}_{{}_A}$ at the output of the EEG decoder is more correlated with the (demixed or microphone) envelope that is associated with the attended speech envelope than with the other envelope, we consider the decoding successful. Here, we consider a (demixed or microphone) envelope to be associated to the attended speech envelope $\mathbf{s}_{{}_A}$ if it has a higher correlation with the attended speech envelope than with the unattended speech envelope. 

We evaluate the performance of the MWF by means of the improvement in the signal-to-noise ratio (SNR). For the different setups of speech sources, we compare the SNR in the microphone with the highest input SNR to the SNR of the output signal of the MWF, i.e.
\begin{align}
\label{eq:MWFSNRin}
\text{SNR}_{\text{in}} &= \underset{i}{\text{max}}\left\{ \frac{\norm{\mathbf{x}_{i1}}^2_2}{\norm{\mathbf{x}_{i2}+\mathbf{v}_i}^2_2}\right\}\\[8pt]
\label{eq:MWFSNRout}
\text{SNR}_{\text{out}}&= \frac{\norm{\sum_{i=1}^{M}\mathbf{w}_i\ast\mathbf{x}_{i1}}^2_2}{\norm{\sum_{i=1}^{M}\mathbf{w}_i\ast\left(\mathbf{x}_{i2}+\mathbf{v}_i\right)}^2_2}
\end{align} %\vphantom{\Bigg()}
where the samples of the signal and noise contributions $x_{i1}[n]$, $x_{i2}[n]$, and $v_i[n]$ from (\ref{eq:convolmix}) are stacked in vectors $\mathbf{x}_{i1}$, $\mathbf{x}_{i2}$, and $\mathbf{v}_{i}$, respectively, covering the full recording length, and $\mathbf{w}_i$ is the time-domain representation of the MWF weights for microphone $i$ (where the WOLA procedure implicitly computes the convolution in (\ref{eq:MWFSNRout}) in the frequency domain). Note that we again assume that $\mathbf{s}_{1}$ represents the attended speech source and $\mathbf{s}_{2}$ is the interfering speech source, which is why $\mathbf{x}_{i2}$ is included in the denominator of (\ref{eq:MWFSNRin}) and (\ref{eq:MWFSNRout}) as it contributes to the (undesired) noise power. Since an unequal number of speaker setups were analyzed at every angular separation, we will mostly consider median SNR values.

\vspace{-4pt}

\subsection{Speech envelope demixing}
\begin{figure}
\centering
\includegraphics[width=.5\textwidth]{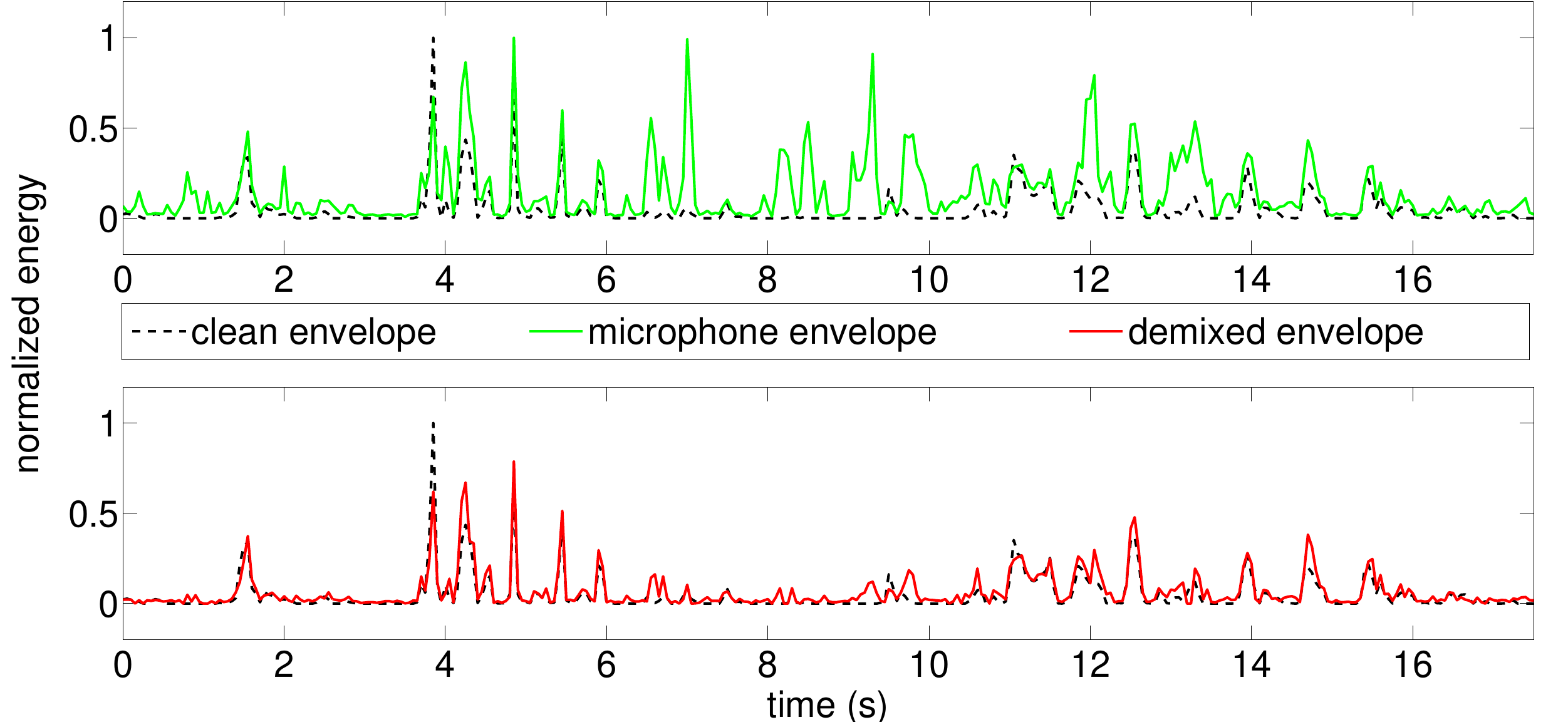}
\caption{Effect of M-NICA, shown for a certain time window. Top figure: original speech envelope (black) and microphone envelope (green). Bottom figure: original speech envelope (black) and demixed envelope (red).}
\label{fig:MNICAenvelopes}
\end{figure}

\begin{figure}
\centering
\includegraphics[width=.5\textwidth]{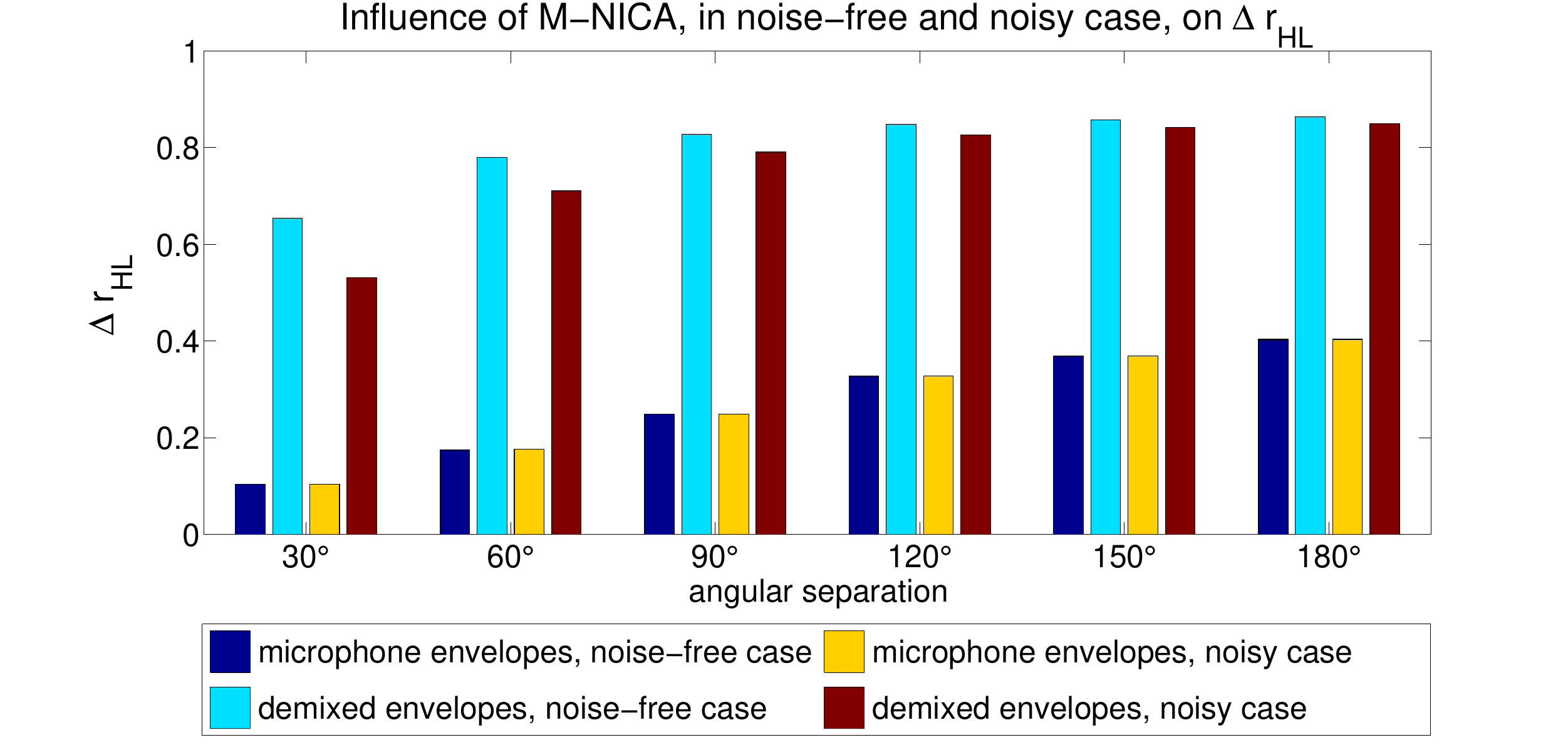}
\caption{Effect of M-NICA: $\Delta r_{{}_{HL}}$ for different separation between the speech sources, for microphone and demixed envelopes in the noise-free case (dark and light blue, respectively) and microphone and demixed envelopes in the noisy case (yellow and red, respectively).}
\label{fig:deltarHL}
\end{figure}

To illustrate the merit of M-NICA as a source separation technique, we plot the different kinds of envelopes in \figurename~\ref{fig:MNICAenvelopes}. In the top figure, the green curve represents an envelope of the speech mixture as observed by a microphone, while the black curve is the envelope of one of the underlying speech sources. The latter is also shown in the bottom figure, together with the corresponding demixed envelope (red curve). All envelopes were rescaled post hoc, because of the ambiguity explained in subsection \ref{subsec:MNICA}. The microphone envelope has spurious bumps, which originate from the energy in the other speech source. The demixed envelope, on the other hand, is a good approximation of the envelope of a single speech source. The improvement of $\Delta r_{{}_{HL}}$ is shown in \figurename~\ref{fig:deltarHL}, for the noise-free and the noisy case. For all relative positions of the speech sources, applying M-NICA to the microphone envelopes gives a substantial improvement in $\Delta r_{{}_{HL}}$, which indicates that the algorithm achieves reasonably good separation of the speech envelopes and hence reduces the crosstalk between them. There is a trend of increasing $\Delta r_{{}_{HL}}$ for speech sources that are wider apart. Indeed, for larger angular separation between the sources, the HRIRs are sufficiently different due to the angle-dependent filtering effects of the head, ensuring energy diversity. The mixing matrix $\mathbf{A}$ will then have weights that make the blind source separation problem defined by (\ref{eq:linearmixing}) better conditioned. When multi-talker background noise is included in the acoustic scene, $\Delta r_{{}_{HL}}$ is seen to be slightly lower, especially for speech sources close together, when the subtle differences in speech attenuation between the microphones are easily masked by noise.

\subsection{AAD performance}
\begin{figure}
\centering
\includegraphics[width=.5\textwidth]{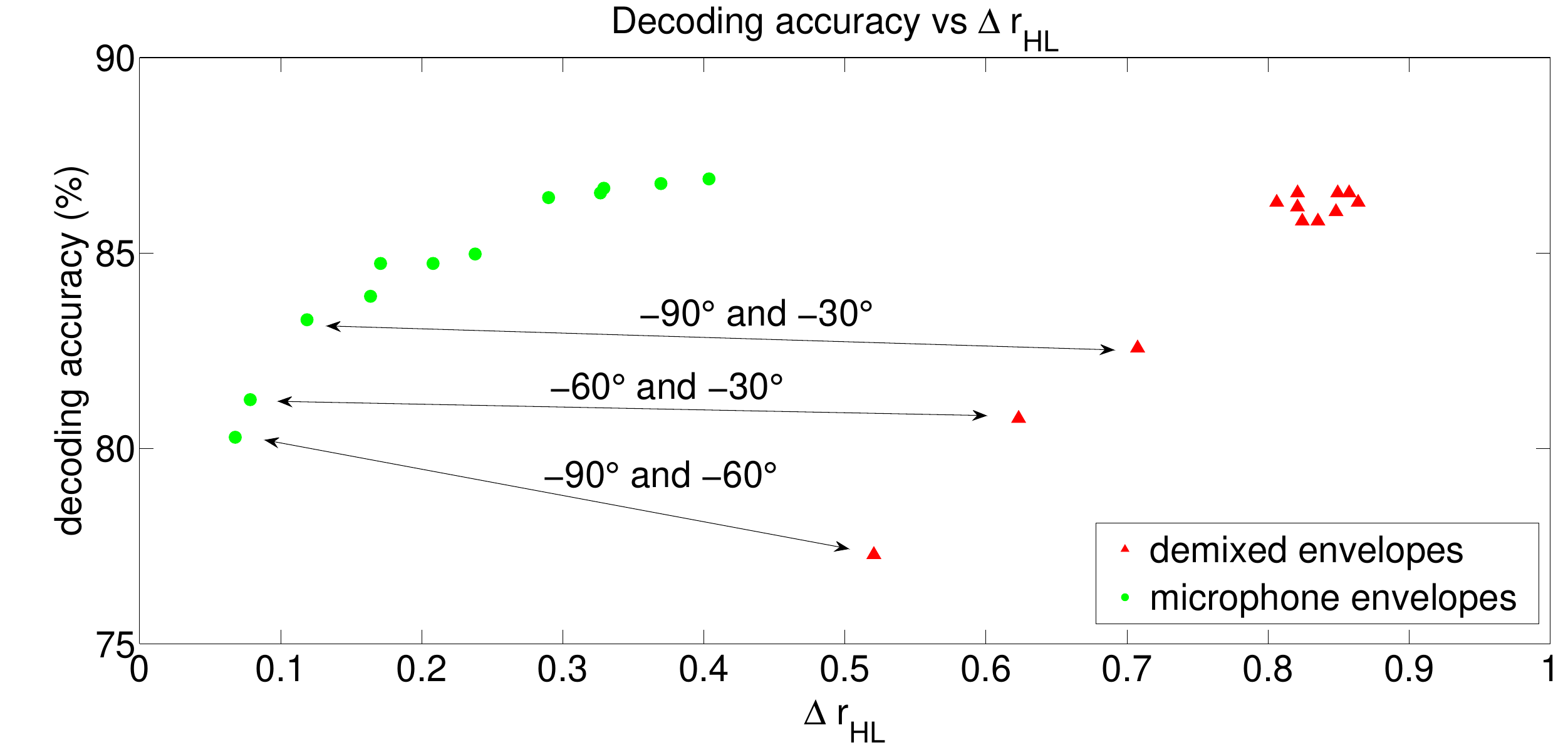}
\caption{Average decoding accuracy over subjects versus $\Delta r_{{}_{HL}}$ for the twelve tested speaker setups, using microphone envelopes (green) or demixed envelopes (red) from the noise-free case. The combinations of speaker positions that lead to the lowest performance are indicated.}
\label{fig:DA}
\end{figure}

\figurename~\ref{fig:DA} shows the average EEG-based AAD accuracy over all subjects versus $\Delta r_{{}_{HL}}$ for different speaker separation angles, when the microphone envelopes or demixed envelopes from the noise-free case are used for AAD. The cluster of points belonging to the demixed envelopes has moved to the right compared to the cluster of the microphone envelopes, conform to what was shown in \figurename~\ref{fig:deltarHL}. Three setups can be distinguished that have a substantially lower AAD accuracy and $\Delta r_{{}_{HL}}$ than the others. Two of them are setups with a separation of $30\,^{\circ}$, while the third one corresponds to a separation of $60\,^{\circ}$. These results are intuitive, as the degree of cross-talk is higher when the speakers are located close to each other. The speakers then have a similar energy contribution to all microphones, which results in lower quality microphone envelopes for AAD and also aggravates the envelope demixing problem, as demonstrated in \figurename~\ref{fig:deltarHL}.

Remarkably, despite the substantial decrease in cross-talk due to the envelope demixing, the average decoding accuracy does not increase when applying the demixing algorithm, i.e., both microphone envelopes and demixed envelopes seem to result in comparable AAD performance.  
However, it is important to put this in perspective, as the accuracy measure for AAD in itself is not perfect (and possibly not entirely representative) when the clean speech signals are not known. Indeed, a `correct' AAD decision here only means that the algorithm selects the candidate envelope that is most correlated to the attended speaker, even if this candidate envelope still contains a lot of crosstalk from the unattended speaker. Therefore, the validity of this measure depends on the quality of the candidate envelopes, i.e., a correct AAD decision according to this principle may have little or no practical relevance if the selected candidate envelope does not contain a high-quality `signature' of the attended speech that can eventually be exploited in the post-processing stage (VAD and MWF) to truly identify or extract the attended speaker. Moreover, M-NICA automatically produces as many candidate envelopes as there are speakers, circumventing the selection of the optimal microphones that would otherwise be necessary, as explained in section \ref{sec:experiment}.

\newpage

\begin{figure}
	\centering
	\subfloat[]{\includegraphics[width=0.9\linewidth]{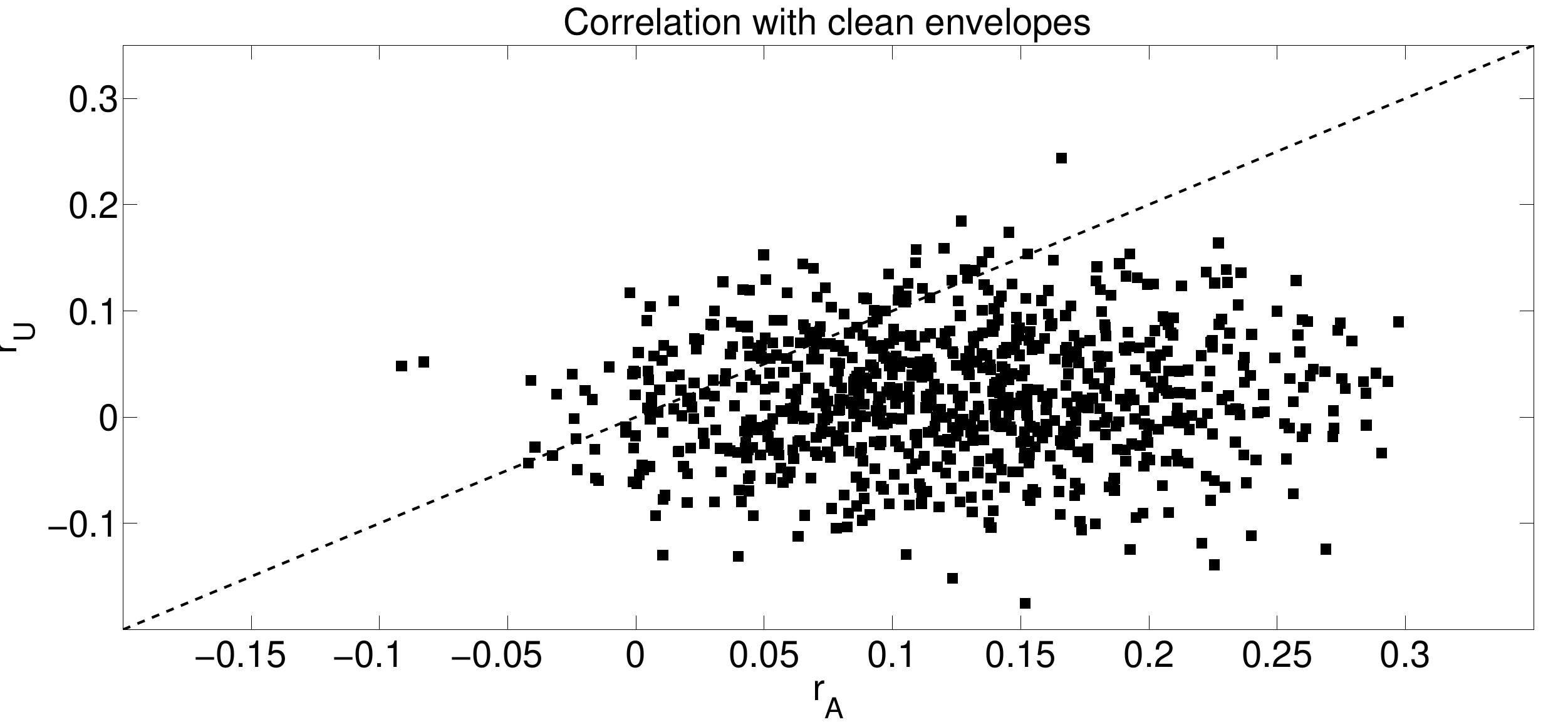}
		\label{fig:scatter_clean}}
	\hfil	
	\subfloat[]{\includegraphics[width=0.9\linewidth]{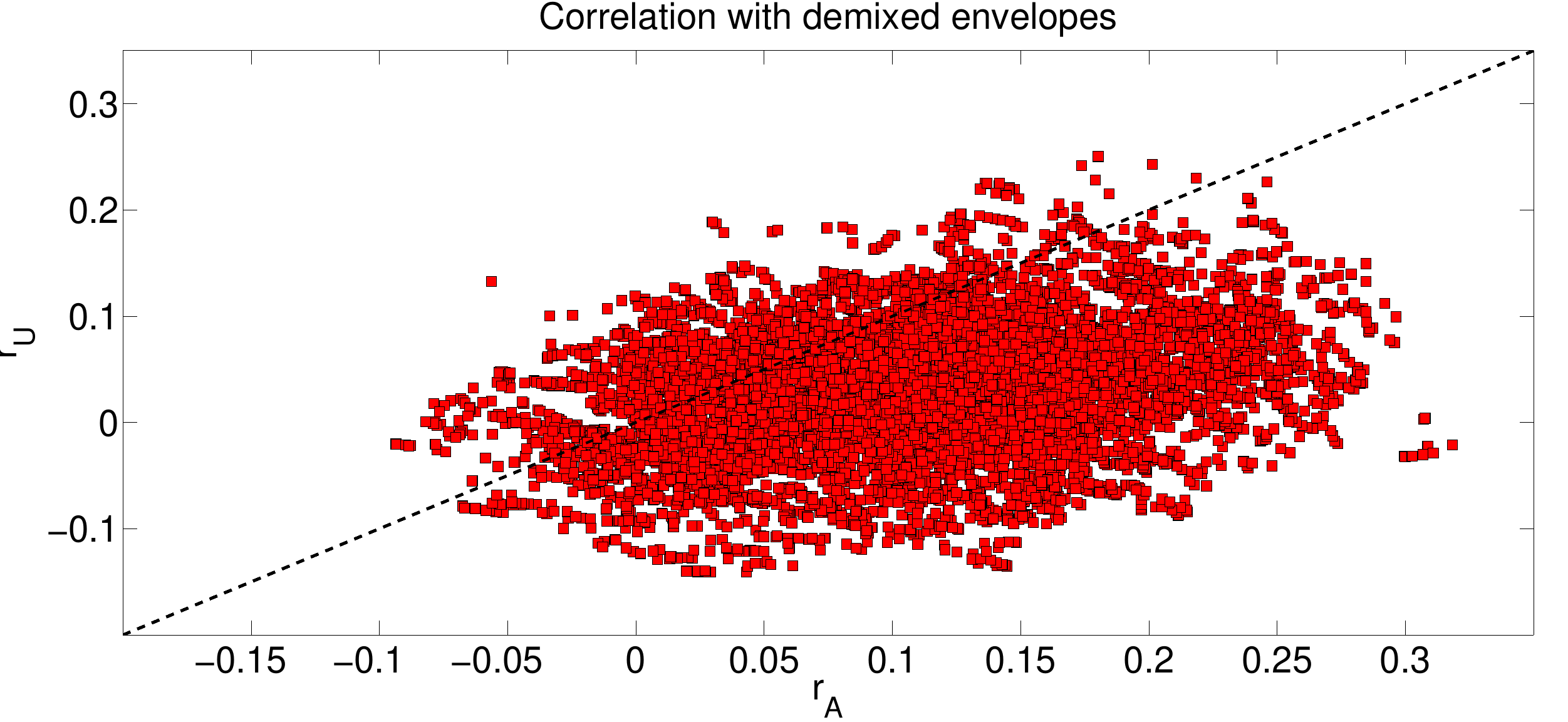}
		\label{fig:scatter_demix}}
	\hfil
	\subfloat[]{\includegraphics[width=0.9\linewidth]{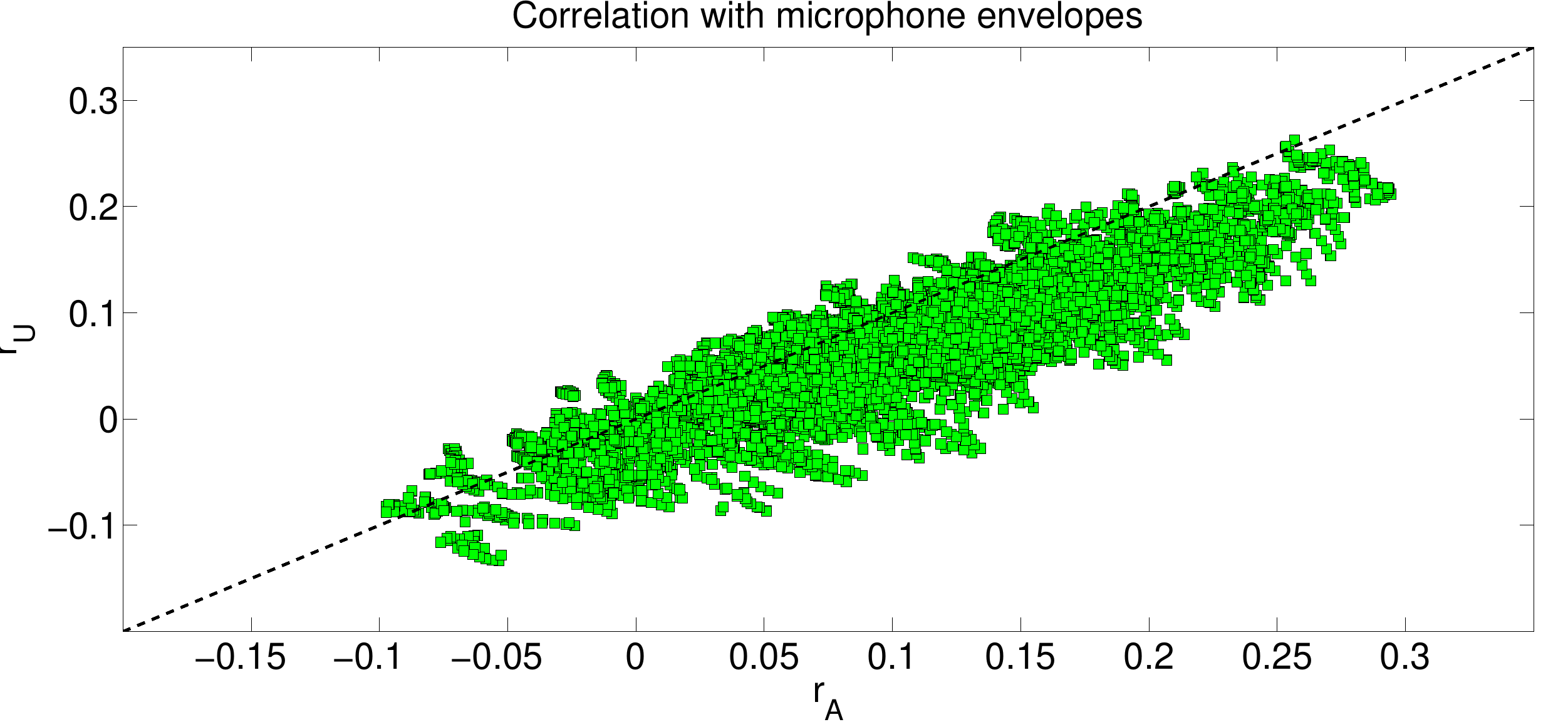}
		\label{fig:scatter_mic}}
	\hfil
	\caption{Scatter plot of the correlation coefficients $r_{{}_{A}}$ and $r_{{}_{U}}$ of the reconstructed envelope with the envelopes of the attended and unattended speech, respectively, for all trials from the noise-free case. Every trial corresponds to one point and is correctly decoded if this point falls below the black decision line $r_{{}_{A}} = r_{{}_{U}}$. The envelopes of the attended and unattended speech are either the clean envelopes (a), demixed envelopes (b), or microphone envelopes (c). Note that the latter two figures consist of more points than the first one, since AAD was performed for 12 different speaker setups.}
	\label{fig:scatter_plots}
\end{figure}

To further illustrate how envelope demixing influences the AAD algorithm, we show in \figurename~\ref{fig:scatter_plots} the correlation of the EEG decoder's output $\mathbf{\widehat{s}}_{{}_A}$ with the true envelopes (in \figurename~\ref{fig:scatter_clean}), and with the two candidate demixed envelopes (in \figurename~\ref{fig:scatter_demix}) as well as with the two candidate microphone envelopes (in \figurename~\ref{fig:scatter_mic}). The point cloud when using the demixed envelopes (\figurename~\ref{fig:scatter_demix}) better resembles the point cloud based on the clean speech envelopes, showing the influence of the demixing process. However, it seems that the variance is higher, as the demixing is not perfect. We observe that the point cloud corresponding to the microphone envelopes (\figurename~\ref{fig:scatter_mic}) is clustered around the main diagonal. Intuitively, this is explained by the fact that the microphone envelopes are not yet separated into separate speech envelopes, and hence they have a considerable mutual resemblance. 

\begin{figure}
\centering
\includegraphics[width=.5\textwidth]{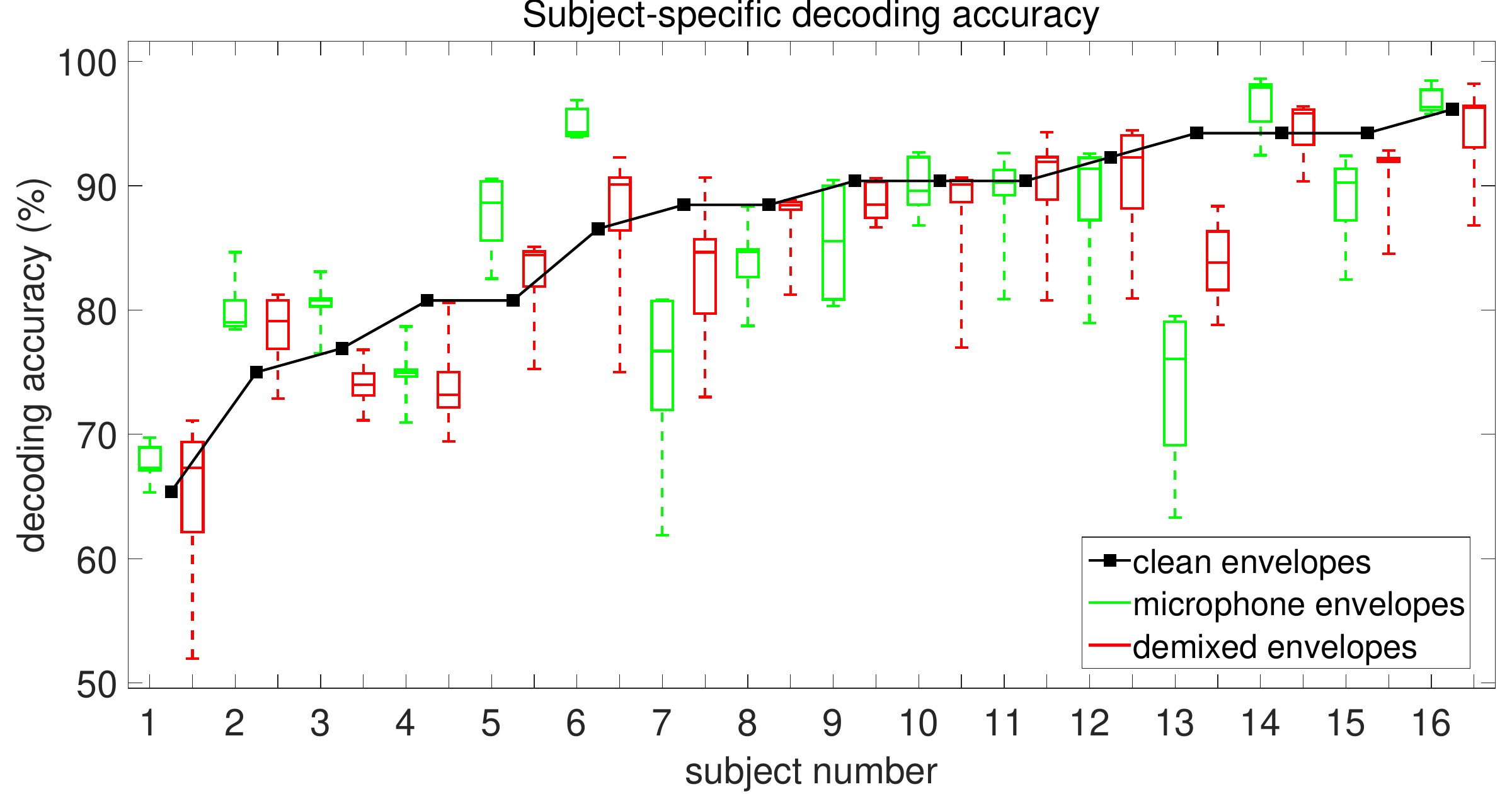}
\caption{Subject-specific decoding accuracy using the accuracy with clean envelopes (black line) as a reference. Accuracies obtained by using microphone (green boxplots) or demixed (red boxplots) envelopes from the noise-free case are shown, over all 12 speaker setups.}
\label{fig:DA_ss}
\end{figure}

Finally, we note that a large variability exists in the decoding accuracy over all subjects, which is illustrated in \figurename~\ref{fig:DA_ss}. It spans a range between 52\% and 98\%, and provides the only subject-specific effect on the overall performance of our processing scheme. The decoding accuracy using either microphone envelopes or demixed envelopes is in general lower than the performance which is obtained using the clean speech envelopes, in an idealized scenario, as expected. Again, we observe that envelope demixing in general does not improve nor lower the AAD accuracy, even if it raises the $\Delta r_{{}_{HL}}$. However, we restate that the AAD accuracy measure employed here is in itself only partially informative. Indeed, this accuracy measure only quantifies how well the AAD algorithm is able to select the envelope with highest correlation with the attended speaker, but not how well this envelope actually represents the attended speaker. The latter is important to also generate an accurate VAD track that only triggers when the attended speaker is active. For this reason, it is relevant to include the demixing step in the analysis, as we show in the next subsection.

\subsection{Denoising and speech extraction performance}\label{subsec:results-denoising}
\begin{figure}
	\centering
	\includegraphics[width=.5\textwidth]{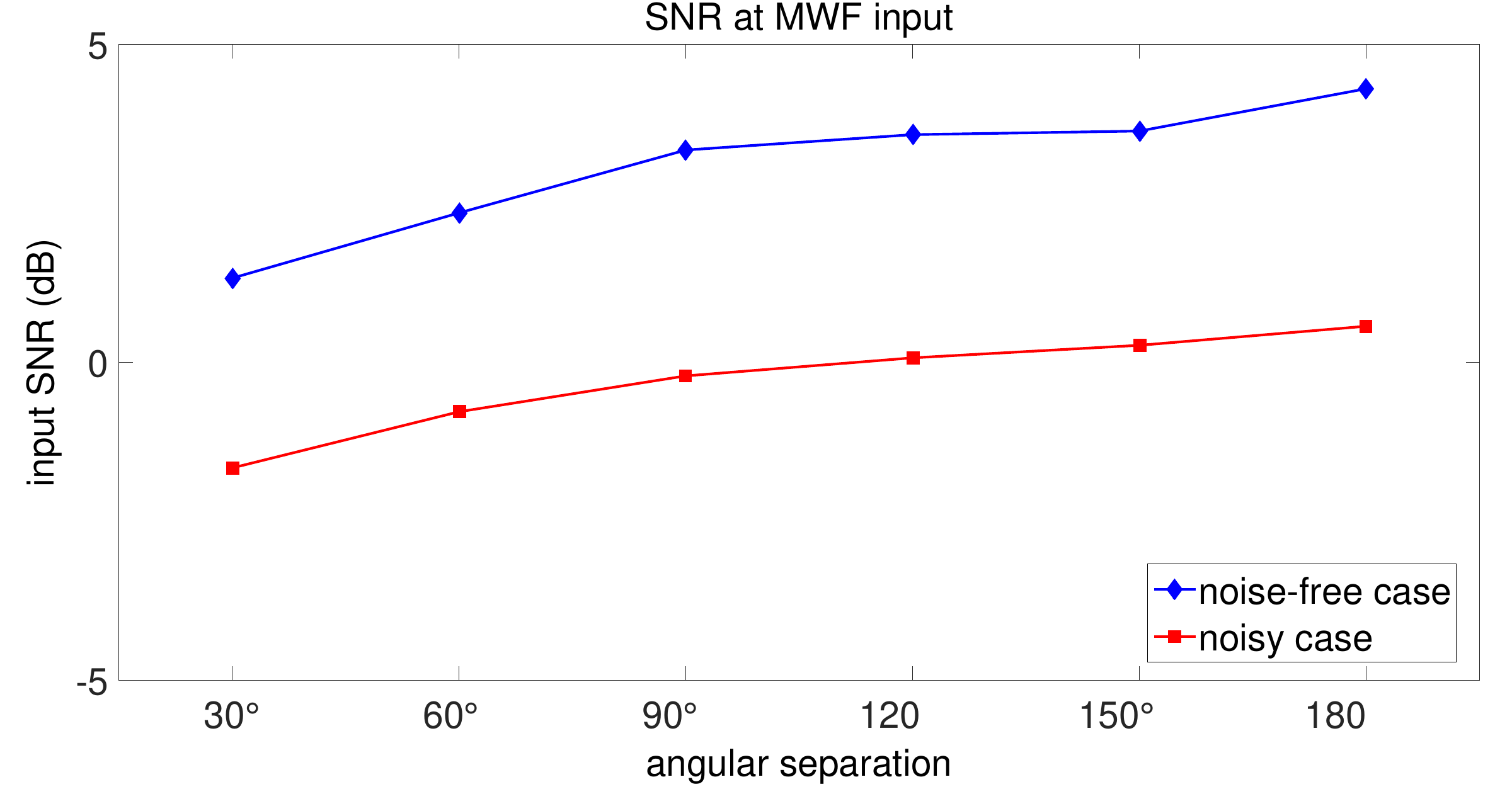}
	\caption{Input SNR taken from the microphone with highest SNR, in the noise-free case (blue) and the noisy case (red), for all angular separations between the speakers.}
	\label{fig:SNRin}
\end{figure}

The median input SNR is shown in \figurename~\ref{fig:SNRin}, for the different angular separations between the speakers, and for both the noise-free and the noisy case. It is noted that in the noisy scenarios, the inclusion of five uncorrelated noise sources with an energy that is 10\% of that of the speech sources, lowers the input SNR with approximately $10\,\text{log}_{10}(5\cdot0.1) = 3$ dB. For equal-energy speech sources that are sufficiently far apart and/or for low noise levels, the input SNR is higher than zero, because in most microphones, one speech source is prevalent over the other due to head shadow and thus for every speech source we can find a microphone signal that gets most of its energy from that particular speech source (recall that the input SNR is defined based on the `best' microphone).

\begin{figure*}
	\centering
	\subfloat[]{\includegraphics[width=0.5\textwidth]{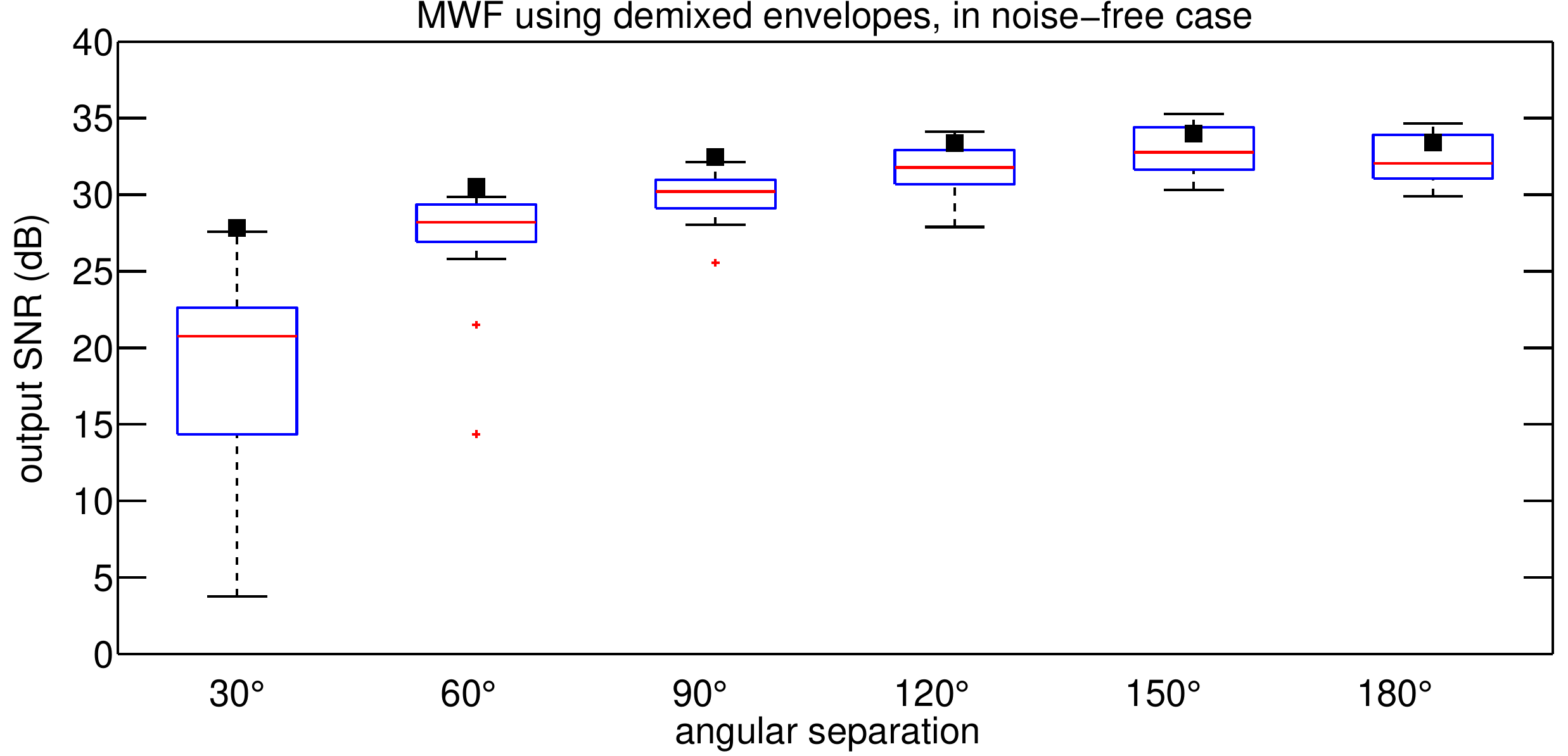}
		\label{fig:SNRdemixclean}}
	\subfloat[]{\includegraphics[width=0.5\textwidth]{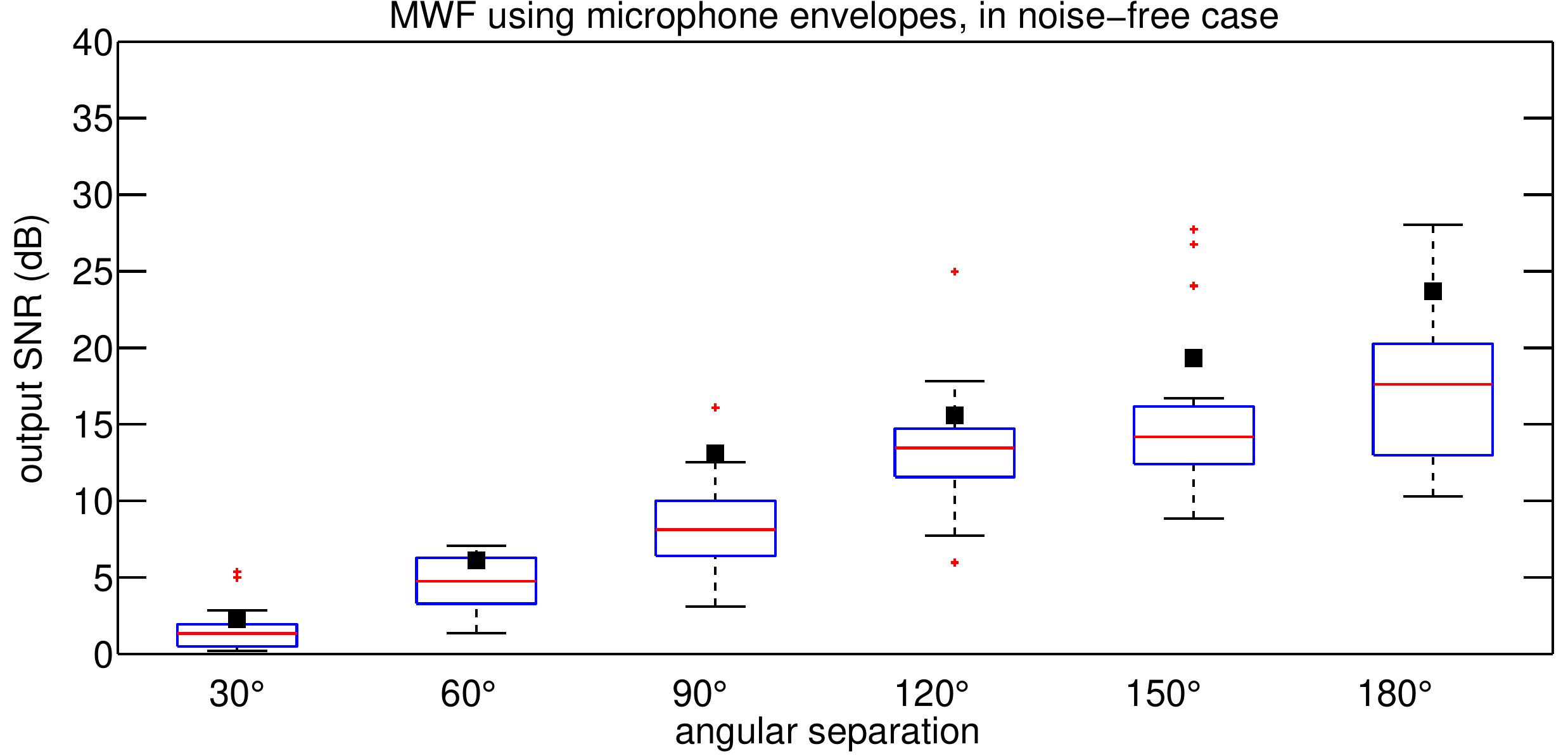}
		\label{fig:SNRmicclean}}
	\hfil
	\subfloat[]{\includegraphics[width=0.5\textwidth]{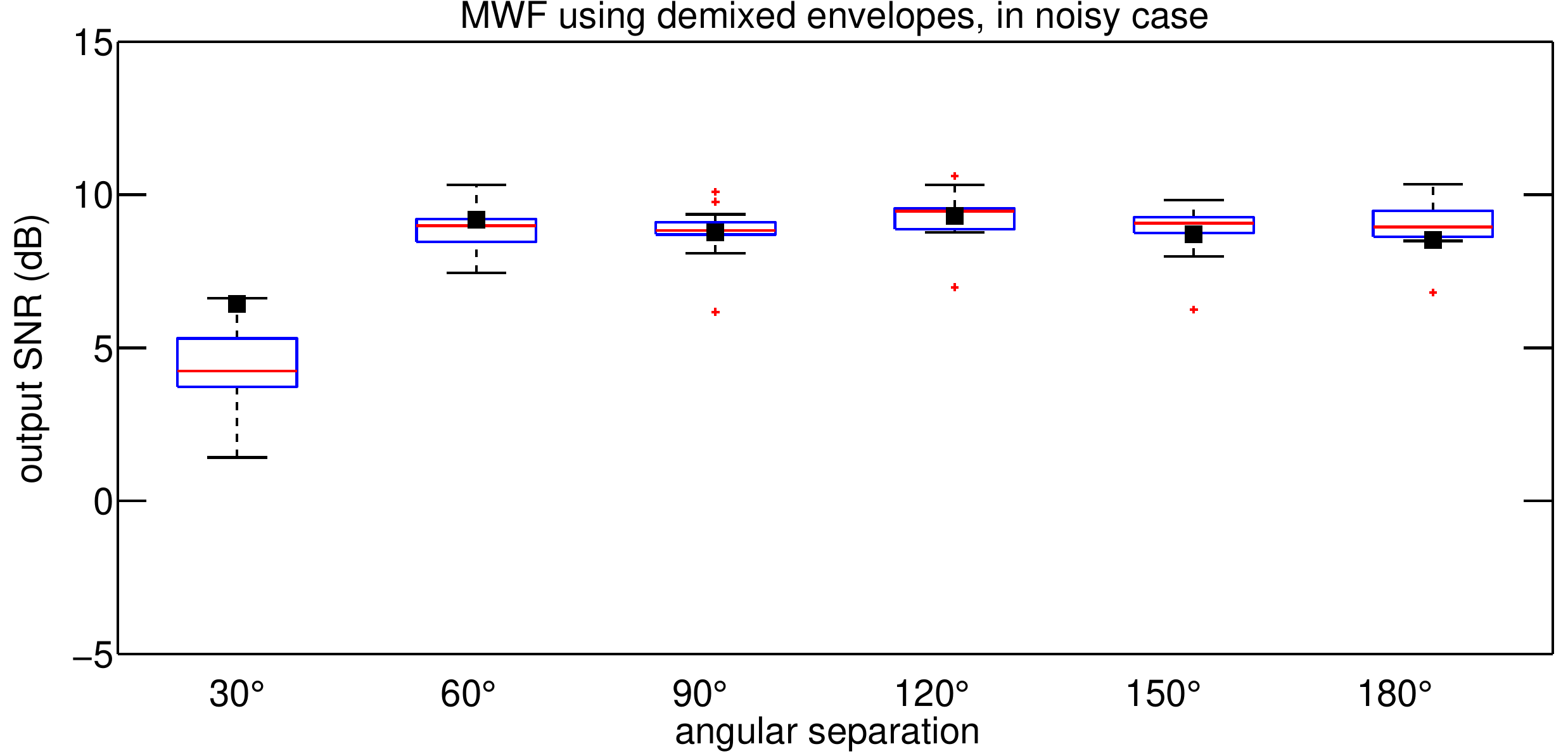}
		\label{fig:SNRdemixBNoise}}
	\subfloat[]{\includegraphics[width=0.5\textwidth]{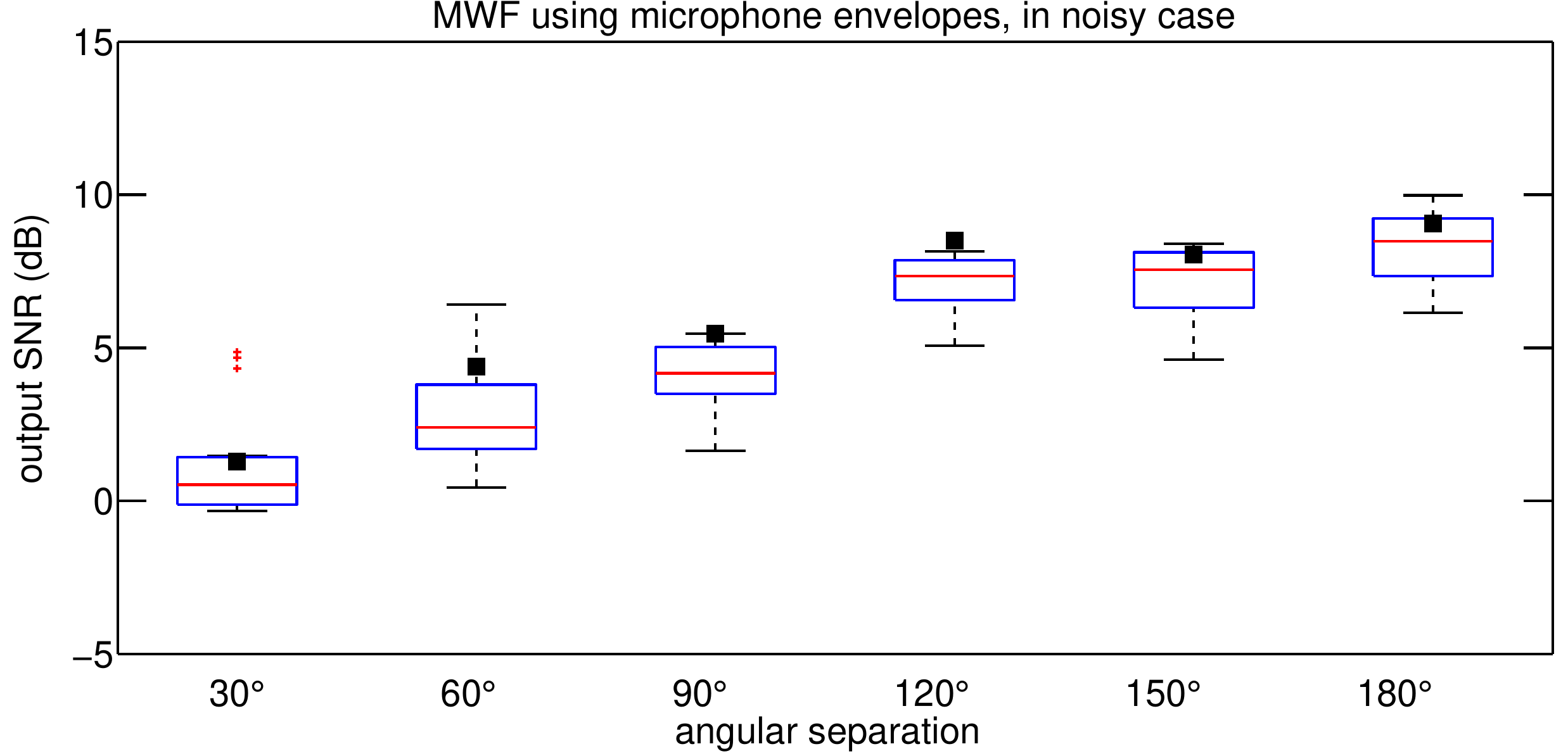}
		\label{fig:SNRmicBNoise}}
	\caption{Boxplots of the output SNR over all subjects and for different angles of speaker separation, using (a) demixed envelopes in the noise-free scenario, (b) microphone envelopes in the noise-free scenario, (c) demixed envelopes in the noisy scenario, (d) microphone envelopes in the noisy scenario. All SNR values represent the median SNR over all pairs of stimuli and possibly multiple speaker setups, per combination of subject and angular separation. The black squares indicate the output SNR for the ideal case of a subject with perfect AAD, i.e. an accuracy of 100\%.}
	\label{fig:SNRMWF}
\end{figure*}

\figurename~\ref{fig:SNRMWF} shows the output SNR for the varying angular separations between the speech sources, ranging from 30$\,^{\circ}$ to 180$\,^{\circ}$. Boxplots show the variation in MWF performance when using the AAD results of each of the 16 subjects (median subject-specific SNR value per angular separation, i.e., 16 values per boxplot). First, we investigate the performance for acoustic setups without additional noise. The output SNR is much higher when computing the AAD/VAD combination based on the demixed envelopes (see \figurename~\ref{fig:SNRdemixclean}), compared to the SNR when computing the AAD/VAD based on the original microphone envelopes (see \figurename~\ref{fig:SNRmicclean}). In the latter case, the performance of the MWF drops as the speech sources are closer together (smaller angular separation). A similar, but smaller effect is observed for the AAD/VAD based on the demixed envelopes. \figurename~\ref{fig:SNRdemixBNoise} and \figurename~\ref{fig:SNRmicBNoise} show the output SNR in the presence of multi-talker background noise when using demixed and microphone envelopes, respectively. In this case, the SNRs are lower - yet still satisfactory, given the sub-zero input SNR - and again the demixed envelopes are seen to be the preferred choice for use in the VAD.
The improvement in SNR when choosing demixed envelopes for the AAD/VAD over the microphone envelopes is significant, both in the noiseless and in the noisy case (p $< 10^{-8}$, 2-way repeated measures ANOVA).
Note that all variability in the SNR over subjects is purely due to the difference in the decoding accuracy, as explained in the previous subsection. The black square markers in the figures show the output SNR for a virtual subject with a decoding accuracy of 100\%. It is seen that the SNR for subjects with a high decoding accuracy closely approximates this ideal performance, and sometimes even surpass it (as the envelopes used for VAD are still imperfect, this is a stochastic effect). As a measure of robustness, we analyzed over which range of VAD thresholds the results we found are valid. From \figurename~\ref{fig:SNRvsthresh}, we see that the VAD based on demixed envelopes gives rise to a high output SNR over a wide range of thresholds. By contrast, when using the microphone envelopes, a low SNR is observed for all thresholds.
The VAD thresholds to generate the results of \figurename~\ref{fig:SNRMWF} were chosen as the optimal values found with these curves, and were reported in subsection \ref{subsec:params}.

\section{Discussion}\label{sec:discussion}
The difference between the SNR at the input and output of the MWF is substantial, demonstrating that MWF denoising can rely on EEG-based auditory attention detection to extract the attended speaker from a set of microphone signals. Furthermore, for the first time, the AAD problem is tackled without use of the clean speech envelopes, i.e., we only use speech mixtures as collected by the microphones of a binaural hearing prosthesis. This serves as a first proof of concept for EEG-informed noise reduction in neuro-steered hearing prostheses.

\begin{figure}[h]
\centering
\includegraphics[width=.5\textwidth]{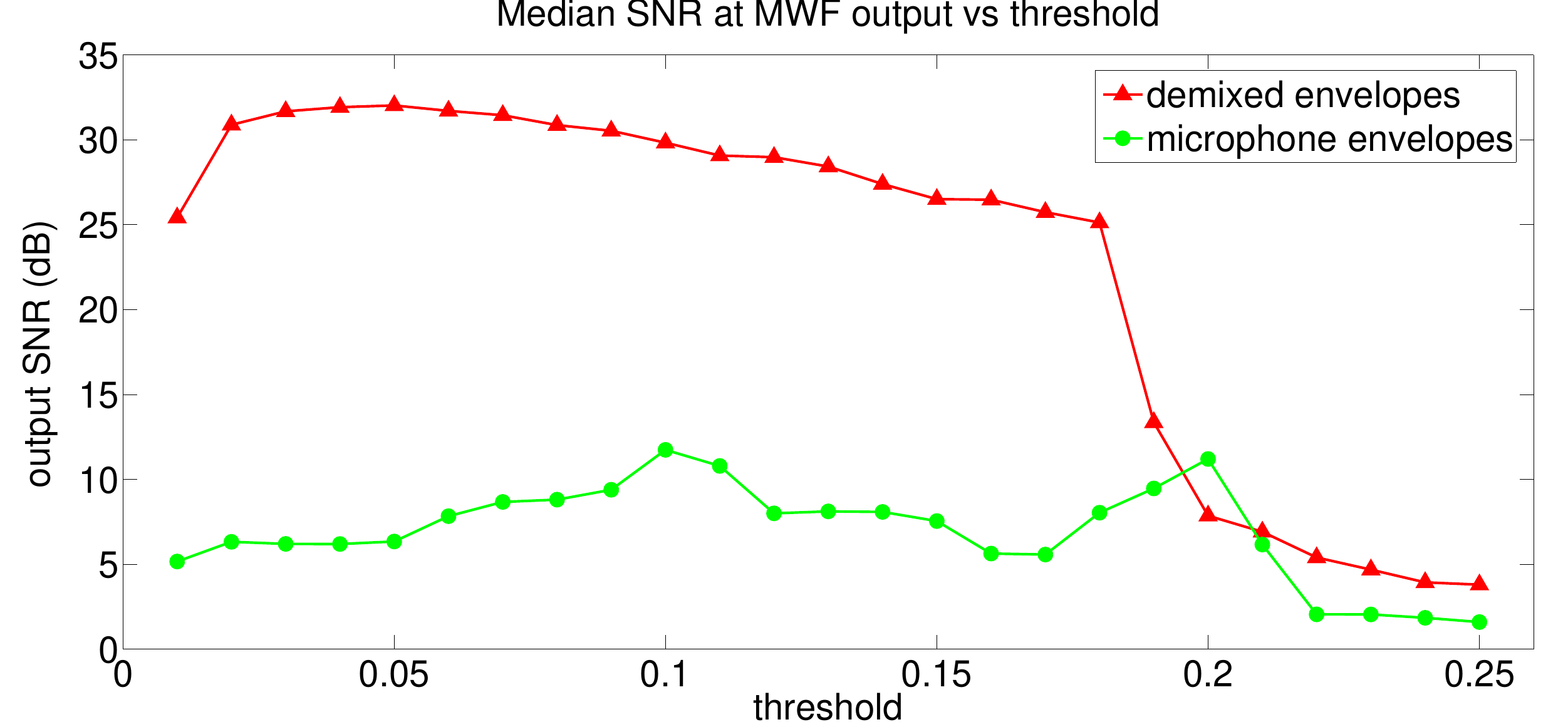}
\caption{SNR at output of the MWF for thresholds going from 1\% to 25\% of the maximum short-term energy, using demixed (red) or microphone (green) envelopes. No multi-talker noise was added, and an idealized AAD track with accuracy of 100\% was used. SNRs are given as the median value over all subjects and all angular separations between the speakers.}
\label{fig:SNRvsthresh}
\end{figure}

Even in severe, noisy environments, subzero input SNRs are boosted to acceptable levels. This positive effect is significantly lower when leaving out the envelope demixing step, showing the necessity of source separation techniques. Rather than applying expensive convolutive ICA methods on the high-rate microphone signals based on higher-order statistics the M-NICA algorithm operates in the low-rate energy domain and only exploits second-order statistics, which makes it computationally attractive. In fact, we circumvent an expensive BSS step on the raw microphone signals by using the fast envelope processing steps and that way postpone the spatiotemporal filtering of the set of microphone signals until the multi-channel Wiener filter. As opposed to convolutive ICA methods, the MWF only extracts a single speaker from a noise background with much lower computational complexity and a higher robustness to noise. From the results in \figurename~\ref{fig:SNRMWF}, we see that the demixing using M-NICA has a strong positive effect on the denoising performance. Although \mbox{M-NICA} indeed slightly improves the AAD accuracy, the use of microphone envelopes without demixing still yields a comparable performance, which is remarkable. The main reason for this is that we always compare with microphones which already have a high $\Delta r_{{}_{HL}}$, i.e., microphones in which one of the two speech sources is already dominant. Such microphone envelopes with sufficiently low crosstalk - resulting in an acceptable AAD accuracy - are present due to the angle-dependent attenuation through the head. In practice however, we do not know which of the microphones provide these good envelopes, which means that the use of M-NICA is still important to obtain a good AAD performance, as it requires no microphone selection. Furthermore, based on \figurename~\ref{fig:SNRvsthresh}, M-NICA seems to lead to more robust VAD results by providing better estimates for the speakers' envelopes, which seems to be the main reason for the improved output SNR when using the MWF. 

The performance of our algorithm pipeline is seen to be robust to the relative speaker position, i.e., even for speakers that are close together, the combination of envelope demixing and multi-channel Wiener filtering results in satisfactory speaker extraction and denoising. The simple VAD scheme proved to be effective, and is insensitive to its threshold setting over a wide range. Note that a straightforward envelope calculation was used for AAD, and that more advanced methods for envelope calculation \cite{biesmansauditory} or for increased robustness in attention detection \cite{akram2016robust} may further increase the accuracy. Also increasing the window length (larger than 30s) improves AAD accuracy, at the cost of a poorer time resolution (the latter is also improved upon in \cite{akram2016robust}). The MWF performance in the case of a perfectly working AAD (shown in \figurename~\ref{fig:SNRMWF}) leads us to believe in the capabilities of the proposed processing flow, especially after incorporation of expected advances in AAD methods.

Future research should aim at collecting EEG measurements from noisy, multi-speaker scenarios over different angles to validate the proposed processing for both the AAD and the speech enhancement on a unified dataset. It should be investigated whether representative EEG can be collected in real life using miniature and semi-invisible EEG devices, e.g., based on in-the-ear \cite{looney2012ear} or around-the-ear EEG \cite{bleichner2015exploring}, and possibly combining multiple such devices \cite{bertrand2014distributed}. A study in \cite{mirkovic2015decoding} has demonstrated that a high AAD accuracy can still be obtained with only 15 EEG channels, although this study assumed availability of the clean speech signals. It has to be investigated whether these results still hold in the case where only the speech mixtures are available, as in this paper. 

As a next step, we aim to adjust the proposed processing scheme to an adaptive implementation, which would be suitable for online, real-time applications. 

\section{Conclusion}\label{sec:conclusion}
We have shown that our proposed algorithm pipeline for EEG-informed speech enhancement or denoising yields promising results in a two-speaker environment, even in conditions with substantial levels of noise. Our technique is extensible to multi-speaker scenarios, and except for an initial training phase, the algorithm operates solely on the microphone recordings of a hearing prosthesis, i.e., without knowledge of the clean speech sources. We have demonstrated that, although the AAD performance decreases, the AAD-informed MWF is still able to extract and denoise the attended speaker with a satisfactory output SNR. All of the elementary building blocks, performing speech envelope demixing, voice activity detection, speech filtering, and auditory attention detection, are computationally inexpensive and are implementable in real-time. This renders them very attractive for use in battery-powered hearing prostheses which have severe constraints on energy usage. With this study, we made the first attempt to bridge the gap between auditory attention detection in ideal scenarios with access to clean speech envelopes, and neuro-steered attended speech enhancement in situations that are more representative for real life environments (without access to the clean speech envelopes). 
\section*{Acknowledgements}

The authors would like to thank Neetha Das and Wouter Biesmans for providing the experimental EEG data and their help with the implementation of the AAD algorithm, and Joseph Szurley for the help with the implementation of the MWF. 

\bibliographystyle{IEEEtran}
\bibliography{bibliografie_morerecent}

\begin{thebibliography}{10}
\providecommand{\url}[1]{#1}
\csname url@samestyle\endcsname
\providecommand{\newblock}{\relax}
\providecommand{\bibinfo}[2]{#2}
\providecommand{\BIBentrySTDinterwordspacing}{\spaceskip=0pt\relax}
\providecommand{\BIBentryALTinterwordstretchfactor}{4}
\providecommand{\BIBentryALTinterwordspacing}{\spaceskip=\fontdimen2\font plus
\BIBentryALTinterwordstretchfactor\fontdimen3\font minus
  \fontdimen4\font\relax}
\providecommand{\BIBforeignlanguage}[2]{{%
\expandafter\ifx\csname l@#1\endcsname\relax
\typeout{** WARNING: IEEEtran.bst: No hyphenation pattern has been}%
\typeout{** loaded for the language `#1'. Using the pattern for}%
\typeout{** the default language instead.}%
\else
\language=\csname l@#1\endcsname
\fi
#2}}
\providecommand{\BIBdecl}{\relax}
\BIBdecl

\bibitem{dillon2001hearing}
H.~Dillon, \emph{Hearing aids}.\hskip 1em plus 0.5em minus 0.4em\relax Thieme,
  2001.

\bibitem{doclo2002gsvd}
S.~Doclo and M.~Moonen, ``{GSVD}-based optimal filtering for single and
  multimicrophone speech enhancement,'' \emph{Signal Processing, IEEE
  Transactions on}, vol.~50, no.~9, pp. 2230--2244, 2002.

\bibitem{serizel2014low}
R.~Serizel \emph{et~al.}, ``Low-rank approximation based multichannel {Wiener}
  filter algorithms for noise reduction with application in cochlear
  implants,'' \emph{Audio, Speech, and Language Processing, IEEE/ACM
  Transactions on}, vol.~22, no.~4, pp. 785--799, 2014.

\bibitem{doclo2009reduced}
S.~Doclo \emph{et~al.}, ``Reduced-bandwidth and distributed {MWF}-based noise
  reduction algorithms for binaural hearing aids,'' \emph{Audio, Speech, and
  Language Processing, IEEE Transactions on}, vol.~17, no.~1, pp. 38--51, 2009.

\bibitem{o2014attentional}
J.~A. O'Sullivan \emph{et~al.}, ``Attentional selection in a cocktail party
  environment can be decoded from single-trial {EEG},'' \emph{Cerebral Cortex},
  p. bht355, 2014.

\bibitem{ding2012emergence}
N.~Ding and J.~Z. Simon, ``Emergence of neural encoding of auditory objects
  while listening to competing speakers,'' \emph{Proceedings of the National
  Academy of Sciences}, vol. 109, no.~29, pp. 11\,854--11\,859, 2012.

\bibitem{golumbic2013mechanisms}
E.~M.~Z. Golumbic \emph{et~al.}, ``Mechanisms underlying selective neuronal
  tracking of attended speech at a ``cocktail party",'' \emph{Neuron}, vol.~77,
  no.~5, pp. 980--991, 2013.

\bibitem{mesgarani2012selective}
N.~Mesgarani and E.~F. Chang, ``Selective cortical representation of attended
  speaker in multi-talker speech perception,'' \emph{Nature}, vol. 485, no.
  7397, pp. 233--236, 2012.

\bibitem{biesmansauditory}
W.~Biesmans \emph{et~al.}, ``Auditory-inspired speech envelope extraction
  methods for improved {EEG}-based auditory attention detection in a cocktail
  party scenario,'' \emph{IEEE Transactions on Neural Systems and
  Rehabilitation Engineering}, vol.~PP, no.~99, pp. 1--1, 2016.

\bibitem{mirkovic2015decoding}
B.~Mirkovic \emph{et~al.}, ``Decoding the attended speech stream with
  multi-channel {EEG}: implications for online, daily-life applications,''
  \emph{Journal of neural engineering}, vol.~12, no.~4, p. 046007, 2015.

\bibitem{aroudiaad}
A.~Aroudi \emph{et~al.}, ``Auditory attention decoding with {EEG} recordings
  using noisy acoustic reference signals,'' in \emph{Acoustics, Speech and
  Signal Processing (ICASSP), 2016 IEEE International Conference on}.\hskip 1em
  plus 0.5em minus 0.4em\relax IEEE, 2016.

\bibitem{mihajlovic2015wearable}
V.~Mihajlovic \emph{et~al.}, ``Wearable, wireless {EEG} solutions in daily life
  applications: what are we missing?'' \emph{Biomedical and Health Informatics,
  IEEE Journal of}, vol.~19, no.~1, pp. 6--21, 2015.

\bibitem{looney2012ear}
D.~Looney \emph{et~al.}, ``The in-the-ear recording concept: User-centered and
  wearable brain monitoring,'' \emph{Pulse, IEEE}, vol.~3, no.~6, pp. 32--42,
  2012.

\bibitem{bleichner2015exploring}
M.~G. Bleichner \emph{et~al.}, ``Exploring miniaturized {EEG} electrodes for
  brain-computer interfaces. an {EEG} you do not see?'' \emph{Physiological
  reports}, vol.~3, no.~4, p. e12362, 2015.

\bibitem{norton2015soft}
J.~J. Norton \emph{et~al.}, ``Soft, curved electrode systems capable of
  integration on the auricle as a persistent brain--computer interface,''
  \emph{Proceedings of the National Academy of Sciences}, vol. 112, no.~13, pp.
  3920--3925, 2015.

\bibitem{bertrand2014distributed}
A.~Bertrand, ``Distributed signal processing for wireless {EEG} sensor
  networks,'' \emph{Neural Systems and Rehabilitation Engineering, IEEE
  Transactions on}, vol.~23, no.~6, pp. 923--935, Nov 2015.

\bibitem{casson2010wearable}
A.~J. Casson \emph{et~al.}, ``Wearable electroencephalography,''
  \emph{Engineering in Medicine and Biology Magazine, IEEE}, vol.~29, no.~3,
  pp. 44--56, 2010.

\bibitem{kayser2009database}
H.~Kayser \emph{et~al.}, ``Database of multichannel in-ear and behind-the-ear
  head-related and binaural room impulse responses,'' \emph{EURASIP Journal on
  Advances in Signal Processing}, vol. 2009, p.~6, 2009.

\bibitem{MNICAconf}
A.~Bertrand and M.~Moonen, ``Energy-based multi-speaker voice activity
  detection with an ad hoc microphone array,'' in \emph{Proc. IEEE Int. Conf.
  Acoustics, Speech, and Signal Processing (ICASSP)}, Dallas, Texas USA, March
  2010, pp. 85--88.

\bibitem{aiken2008human}
S.~J. Aiken and T.~W. Picton, ``Human cortical responses to the speech
  envelope,'' \emph{Ear and hearing}, vol.~29, no.~2, pp. 139--157, 2008.

\bibitem{pasley2012reconstructing}
B.~N. Pasley \emph{et~al.}, ``Reconstructing speech from human auditory
  cortex,'' \emph{PLoS-Biology}, vol.~10, no.~1, p. 175, 2012.

\bibitem{ding2012neural}
N.~Ding and J.~Z. Simon, ``Neural coding of continuous speech in auditory
  cortex during monaural and dichotic listening,'' \emph{Journal of
  neurophysiology}, vol. 107, no.~1, pp. 78--89, 2012.

\bibitem{kerlin2010attentional}
J.~R. Kerlin \emph{et~al.}, ``Attentional gain control of ongoing cortical
  speech representations in a {``}cocktail party{''},'' \emph{The Journal of
  Neuroscience}, vol.~30, no.~2, pp. 620--628, 2010.

\bibitem{bertrand2010blind}
A.~Bertrand and M.~Moonen, ``Blind separation of non-negative source signals
  using multiplicative updates and subspace projection,'' \emph{Signal
  Processing}, vol.~90, no.~10, pp. 2877--2890, 2010.

\bibitem{chouvardas2015distributed}
S.~Chouvardas \emph{et~al.}, ``Distributed robust labeling of audio sources in
  heterogeneous wireless sensor networks,'' in \emph{Acoustics, Speech and
  Signal Processing (ICASSP), 2015 IEEE International Conference on}.\hskip 1em
  plus 0.5em minus 0.4em\relax IEEE, 2015, pp. 5783--5787.

\bibitem{BertrandIWAENC10}
A.~Bertrand \emph{et~al.}, ``Adaptive distributed noise reduction for speech
  enhancement in wireless acoustic sensor networks,'' in \emph{Proc. of the
  International Workshop on Acoustic Echo and Noise Control (IWAENC)}, Tel
  Aviv, Israel, August 2010.

\bibitem{akram2016robust}
S.~Akram \emph{et~al.}, ``Robust decoding of selective auditory attention from
  meg in a competing-speaker environment via state-space modeling,''
  \emph{NeuroImage}, vol. 124, pp. 906--917, 2016.

\end{thebibliography}
\end{document}